\documentclass[english]{revtex4}
\usepackage[T1]{fontenc}
\usepackage[latin9]{inputenc}
\usepackage{float}
\usepackage{amsmath}
\usepackage{amssymb}

\makeatletter

\providecommand{\tabularnewline}{\\}

\@ifundefined{textcolor}{}
{%
 \definecolor{BLACK}{gray}{0}
 \definecolor{WHITE}{gray}{1}
 \definecolor{RED}{rgb}{1,0,0}
 \definecolor{GREEN}{rgb}{0,1,0}
 \definecolor{BLUE}{rgb}{0,0,1}
 \definecolor{CYAN}{cmyk}{1,0,0,0}
 \definecolor{MAGENTA}{cmyk}{0,1,0,0}
 \definecolor{YELLOW}{cmyk}{0,0,1,0}
}

\makeatother

\usepackage{babel}
\begin{document}
\title{A Statistical Fields Theory underlying the Thermodynamics of Ricci
Flow and Gravity}
\author{M.J.Luo}
\address{Department of Physics, Jiangsu University, Zhenjiang 212013, People's
Republic of China}
\email{mjluo@ujs.edu.cn}

\begin{abstract}
The paper proposes a statistical fields theory of quantum reference
frame underlying the Perelman's analogies between his formalism of
the Ricci flow and the thermodynamics. The theory is based on a $d=4-\epsilon$
quantum non-linear sigma model (NLSM), interpreted as a quantum reference
frame system which a to-be-studied quantum system is relative to.
The statistic physics and thermodynamics of the quantum frame fields
is studied by the density matrix obtained by the Gaussian approximation
quantization. The induced Ricci flow of the frame fields and the Ricci-DeTurck
flow of the frame fields associated with the density matrix is deduced.
In this framework, the diffeomorphism anomaly of the theory has a
deep thermodynamic interpretation. The trace anomaly is related to
a Shannon entropy in terms of the density matrix, which monotonically
flows and achieves its maximal value at the flow limit, called the
Gradient Shrinking Ricci Soliton (GSRS), corresponding to a thermal
equilibrium state of spacetime. A relative Shannon entropy w.r.t.
the maximal entropy gives a statistical interpretation to Perelman's
partition function, which is also monotonic and gives an analogous
H-theorem to the statistical frame fields system. A temporal static
3-space of a GSRS 4-spacetime is also a GSRS in lower 3-dimensional,
we find that it is in a thermal equilibrium state, and Perelman's
analogies between his formalism and the thermodynamics of the frame
fields in equilibrium can be explicitly given in the framework. Extending
the validity of the Equivalence Principle to the quantum level, the
quantum reference frame fields theory at low energy gives an effective
theory of gravity, a scale dependent Einstein-Hilbert action plus
a cosmological constant is recovered. As a possible underlying microscopic
theory of the gravitational system, the theory is also applied to
understand the thermodynamics of the Schwarzschild black hole. 
\end{abstract}
\maketitle

\section{Introduction}

Recent works \citep{Luo:2019iby,Luo:2021zpi} show possible relations
between Perelman's formalism of the Ricci flow and some fundamental
problems in quantum spacetime and quantum gravity, for instance, the
trace anomaly and the cosmological constant problem. Perelman's seminal
works (the section-5 of \citep{perelman2002entropy}) and further
development by Li \citep{2012Perelman,2013arXiv1303.5193L} also suggest
deep relations between the Ricci flow and the thermodynamics system,
not only the irreversible non-equilibrium but also the thermal equilibrium
thermodynamics of certain underlying microscopic system. In \citep{perelman2002entropy}
Perelman also declared a partition function and his functionals without
specifying what the underlying microscopic ensemble really are (in
physics). So far it is not clear whether the beautiful thermodynamic
analogies are physical or pure coincidences. On the other hand, inspired
by the surprising analogies between the black hole and thermodynamics
system, it is generally believed the existence of temperature and
entropy of a black hole. Works along this line also showed, in many
aspects, the gravitational system would be profoundly related to thermodynamics
system (see recent review \citep{Page:2004xp} and references therein),
it is generally conjectured that there would exist certain underlying
statistical theory for the underlying microscopic quantum degrees
of freedom of gravity. It gradually becomes one of the touchstones
for a quantum gravity. 

The motivations of the paper are, firstly, to propose an underlying
statistical fields theory for Perelman's seminal thermodynamics analogies
of his formalism of the Ricci flow, and secondly, for understanding
the possible microscopic origin of the spacetime thermodynamics especially
for the Schwarzschild black hole. We hope the paper could push forward
the understanding to the possible interplay of the mysterious Perelman's
formalism of Ricci flow and the quantum spacetime and gravity. To
our knowledge, several tentative works have been devoted to the goal,
see e.g. \citep{Shu:2006bk,Headrick:2006ti,Samuel:2007ak,Samuel:2007zz},
but frankly speaking, the physical picture underlying the Ricci flow
is not fully clear, if a fundamental physical theory underlying the
Ricci flow and a fundamental theory of quantum spacetime is lacking.

Based on our previous works \citep{Luo2014The,Luo2015Dark,Luo:2015pca,Luo:2019iby,Luo:2021wdh,Luo:2021zpi,Luo:2022goc,Luo:2022ywl}
on the quantum reference frame and its relation to Perelman's formalism
of the Ricci flow, we propose a statistical fields theory of the quantum
reference frame as a possible underlying theory of Perelman's seminal
analogies between his geometric functionals and the thermodynamic
functions. In section II, we review the theory of quantum reference
frame based on a $d=4-\epsilon$ quantum non-linear sigma model, at
the Gaussian approximation quantization, we obtain a density matrix
of the frame fields system as a physical foundation to the statistical
interpretation of the theory. The induced Ricci flow of the frame
fields and the Ricci-DeTurck flow of the frame fields associated with
the density matrix is deduced. In section III, we discuss the diffeomorphism
and related trace anomaly of the quantum frame fields theory and its
profound implications to the irreversible non-equilibrium thermodynamics
of the frame fields, for instance, the statistical entropy and an
analogous H-theorem of the frame fields, and the effective gravity
theory at cosmic scale (especially the emergence of the cosmological
constant). In section IV, the thermal equilibrium state of the frame
fields as a flow limit configuration (the Gradient Shrinking Ricci
Soliton) is discussed, in which the density matrix recovers the thermal
equilibrium canonical ensemble density. This section gives a physical
foundation to Perelman's seminal thermodynamic analogies. In section
V, the framework gives a possible microscopic understanding of the
thermodynamics of the Schwarzschild black hole. Finally, we summarize
the paper and give conclusions in the section VI.

\section{Quantum Reference Frame}

Reference frame is one of the most fundamental notions in physics.
Any measurement in physics is performed or described, a reference
frame has always been explicitly or implicitly used. In classical
physics, the reference frame is idealizationally used via classical
rulers and clocks to label the spacetime coordinates, which are classical,
external, and rigid without any fluctuation. Even in the textbook
quantum mechanics or quantum fields theory, the spacetime coordinates
are still classical. But quantum principles tell us that all physical
measuring instruments including the rulers and clocks are inescapably
quantum fluctuating. Such idealizational and classical treatment of
reference frame works not bad in quantum mechanics and quantum fields
theory. To a large extent, this is due to the fact that the general
coordinates transformation and gravitational effects are not seriously
taken into account. Just as expected, when the quantum principles
are seriously applied to the spacetime itself and gravitational phenomenon,
severe difficulties arise, e.g. information losses (non-unitary),
diffeomorphism anomaly and the cosmological constant problems, etc.

The quantum reference frame is a recurring theme in literature (not
completely list, see e.g. \citep{1984PhRvD..30..368A,rovelli1991quantum,dickson2004view,2007Reference,2007Fundamental,angelo2011physics,Flaminia2019Quantum,Hoehn:2020epv}
and references therein) based on various difference physical motivations,
from quantum foundation to quantum information or quantum communication,
to quantum gravity. For example, in Ref.\citep{1984PhRvD..30..368A},
the author suggests the general relation between superselection rules
and the lack of reference frame. In Ref.\citep{2007Reference}, it
also more practically shows that extra assumptions about the superselection
rules can not be avoided from the viewpoint of quantum information
and quantum communication theory, if local observers do not share
common information about their relative phase or Cartesian frames
etc. The extra assumptions of the superselection rules may be also
viewed as the weakness of textbook quantum mechanics, which can be
overcome by introducing appropriate quantum reference frame. And many
models (e.g. \citep{rovelli1991quantum,angelo2011physics,Flaminia2019Quantum})
of quantum reference frame and relational descriptions to the quantum
system and the quantum reference frame as a whole are suggest into
the quantum foundation. In recent works \citep{Hoehn:2020epv} and
the references therein, the authors review three approaches (relational
Dirac observables, the Page-Wootters formalism and quantum deparameterizations)
of relational quantum dynamics and suggest their equivalence. Other
author focus on the possible role of quantum reference frame to the
decoherence in quantum gravity \citep{Poulin:2005dn,2007Fundamental}.
Certainly, the works list of the direction is far from complete, which
is beyond the scope and ability of the author. 

Fundamentally, our work shares the similar philosophical viewpoint
to the role of quantum reference frame in quantum mechanics, such
as considering that an appropriate materialized (but idealized) reference
frame obeying the same law of quantum mechanics must be taken into
account, and in the full quantum theory a relational description based
on an entanglement of a quantum system and the quantum reference frame
as a whole must play a fundamental role. However, there are some differences
from the past literature that we considered more important. First,
we do not simply or merely treat the quantum clock as a quantum mechanical
system (\citep{Flaminia2019Quantum,Hoehn:2020epv}) (which is more
simple and has less degree of free to deal with as discussed in most
quantum reference literature, in fact our early work (\citep{Luo2014The,Luo2015Dark})
also started from the operationally treatment of quantum clock to
get some general conclusions on the vaccum energy and the cosmological
constant problem), but in the paper we put both quantum space-rod
and clock-time on an equal footing in the framework of quantum statistical
fields, so that it makes the theory more appropriate to incorporate
gravity, under the assumption of a quantum version of equivalence
principle. To my understanding, quantum clock can be viewed as a first
step model and far from a theory. Second, based on the quantum spacetime
reference frame model (i.e. the $d=4-\epsilon$ non-linear sigma model),
our paper does not treat the genuine relational quantities from the
very beginning (as most literature tend to announce), but we prepare
the quantum frame fields of reference in a laboratory frame (the $d=4-\epsilon$
base spacetime of the non-linear sigma model) as the starting reference,
and then quantum events are relative to the prepared quantum frame
fields. In this sense, the framework equivalently assumes the existence
of an external, classical and rigid (free from quantum fluctuation
and volume fixed) reference frame to be the laboratory frame, since
the non-linear sigma model allows us to assign quantum state of spacetime
reference (the target spacetime) to the base spacetime to arbitrary
precision. But it can be easily verified that the theory is independent
to the laboratory frame (metric, sign etc.) in the non-linear sigma
model. The notion of the external and classical laboratory frame is
just for convenience, since a quantum statistical fields theory is
historically (or maybe more appropriate to) defined on an inertial
frame (flat spacetime). So the relational quantities describing the
relation between the quantum system and the quantum spacetime reference
system is in essential in the framework. Third, also for the reason
of the base spacetime independence of the non-linear sigma model,
whose Hamiltonian is trivial, so the theory of spacetime reference
frame is more properly quantized by using the path integral or functional
method rather than the operator methods (e.g. the relational Dirac
observables quantization or relational Schrodinger picture in Page-Wootters
formalism). And fourth, there is a fundamentally non-unitary relation
between two spacetime reference frames under a coordinate transformation
due to an irreversible Ricci flow of spacetime reference frame, unlike
most approaches in which the coordinate transformation between difference
reference frames is assumed unitary. This is considered as a key ingredient
of quantum spacetime reference frame that is intrinsically ensemble
statistical and thermal.

Generally speaking, our approach alongs the general philosophy of
the quantum reference frame but is considered independent to the details
of the past literature. The framework associates to several elegant
physics and mathematical structures that are not discussed in the
past literature, such as the non-linear sigma model, Shannon entropy,
the Ricci flow and density Riemannian geometry, etc. And our previous
works \citep{Luo2014The,Luo2015Dark,Luo:2015pca,Luo:2019iby,Luo:2021wdh,Luo:2021zpi,Luo:2022goc,Luo:2022ywl}
have revealed very rich consequencies of the framework, (e.g. the
acceleration expansion of the late epoch universe, the cosmological
constant, diffemorphism anomaly, the inflationary early universe,
local conformal stability and non-collapsibility, modified gravity,
etc.), but frankly speaking, the possible consequencies of the quantum
reference frame are still far from fully discovered. The main motivation
here for a quantum treatment of a reference frame system is that it
might form a foundation to constructing a theory of quantum spacetime
and quantum gravity that is analogous to the way it is used to construct
the classical general relativity and it is crucial in understanding
the microscopic origin of the spacetime thermodynamics.

\subsection{Definition}

In this section, we propose a quantum fields theory of reference frame
as a starting point to study a quantum theory of spacetime and quantum
gravity, based on an Equivalence Principle extended to reference frame
described by quantum state (discussed by a paradox in Section-V-B
and in the conclusion of the paper). The generalization of the Equivalence
Principle to the quantum level might form another foundation to a
quantum reference frame and a quantum gravity. How the Equivalence
Principle behaves at the quantum level has many discussions having
a long history (e.g. \citep{1980The,1983Is,1994On,2010arXiv1011.3719G,2019arXiv190301289H}
and references therein, and \citep{2013arXiv1308.6289K,2016arXiv161001968S}
for an extended thermal version). The Equivalence Principle is the
physical foundation of measuring the spacetime by physical material
reference frame even at the quantum level, and it is the bridge between
the geometric curved spacetime and gravity, and hence the gravity
is simply a relational phenomenon that the motion of a test particle
in gravity is manifested as a relative motion w.r.t. the (quantum)
material reference frame. Without the Equivalence Principle, we would
lost the physical foundation of all these concepts. Therefore, the
basic argument of the paper is that there are several supports (e.g.
uniform quantum origin of the accelerating expansion of the universe
posited by myself in previous works \citep{Luo:2015pca,Luo:2019iby,Luo:2021zpi},
and a consistent incorporating the thermodynamics of the spacetime
shown in this work) and the self-consistency of the framework are
all possible evidences for its validity for the quantum reference
frame. 

In this framework, a to-be-studied quantum system described by a state
$|\psi\rangle$ and the spacetime reference system by $|X\rangle$
are both quantum. The states of the whole system are given by an entangled
state
\begin{equation}
|\psi[X]\rangle=\sum_{ij}\alpha_{ij}|\psi\rangle_{i}\otimes|X\rangle_{j}\label{eq:entangle}
\end{equation}
in their direct product Hilbert space $\mathcal{H}_{\psi}\otimes\mathcal{H}_{X}$.
The state (\ref{eq:entangle}) of the to-be-studied system and the
reference frame system is an entangled state but a trivial direct
product state is for the reason of calibration between them. Usually,
a quantum measurement is performed as follows. At a preparation step
of a quantum measurement, a one-to-one correlation between a quantum
system $|\psi\rangle_{i}$ and a reference system $|X\rangle_{j}$
(a quantum instrument or ruler) is prepared, called calibration. The
step in usual sense is a comparison and adjustment of the measuring
instrument $|X\rangle_{j}$ by a calibration standard $|\psi_{standard}\rangle_{i}$
which is physically similar with the to-be-studied system $|\psi\rangle_{i}\doteq|\psi_{standard}\rangle_{i}$.
A well-calibrated entangled state $\sum_{ij}\alpha_{ij}|\psi_{standard}\rangle_{i}\otimes|X\rangle_{j}$
can be used to measure the to-be-studied system $|\psi\rangle_{i}$
with the reference to the quantum instrument $|X\rangle_{j}$. In
essential, the measurement indirectly performs a comparision between
$|\psi\rangle_{i}$ and the fiducial state $|\psi_{standard}\rangle_{i}$.
So the entangled state $|\psi[X]\rangle$ is a superposition of all
possible one-to-one correlations. According to the standard Copenhagen
interpretation of a quantum state, the to-be-studied quantum system
collapsing into a state $|\psi\rangle_{i}$ together with the collapsing
of the quantum reference system into the corresponding $|X\rangle_{j}$
happening by the joint probability $|\alpha_{ij}|^{2}$, meaning that
when the state of the quantum instrument is read out being in state
$|X\rangle_{j}$, then in this sense the to-be-studied system is inferred
to be the corresponding $|\psi\rangle_{i}$. A simple and practical
example is the Stern-Gerlach experiment (see \citep{Luo:2019iby}).
The entangled state generalizes the textbook quantum description of
the state $|\psi(x)\rangle$ w.r.t. an idealized parameter $x$ of
a classical reference system free from quantum fluctuations (in quantum
mechanics $x$ is the Newtonian time, in quantum fields theory $x_{a}$
are the Minkowskian spacetime).

The entangled state $|\psi[X]\rangle$ is inseparable, so that the
state can only be interpreted in a relational manner, i.e. the entangled
state describes the ``relation'' between $|\psi\rangle$ and $|X\rangle$,
but each absolute state. The individual state $|\psi\rangle$ has
physical meaning only being reference to $|X\rangle$ entangled to
it. When quantum mechanics is reformulated on the new foundation of
the relational quantum state (the entangled state) describing the
``relation'' between the state of the under-studied quantum system
and the state of the quantum reference system, a gravitational theory
is automatically contained in the quantum framework without extra
assumption.

Since the state of reference $|X\rangle$ is also subject to quantum
fluctuation, so mathematically speaking, the state $|\psi[X]\rangle$
can be seen as the state $|\psi(x)\rangle$ with a smeared spacetime
coordinates, instead of the textbook state $|\psi(x)\rangle$ with
a definite and classical spacetime coordinates. The state $|\psi[X]\rangle$
could recover the textbook state $|\psi(x)\rangle$ only when the
quantum fluctuation of the reference system is small enough and hence
can be ignored. More precise, the 2nd order central moment (even higher
order central moments) fluctuations of the spacetime coordinate $\langle\delta X^{2}\rangle$
(the variance) can be ignored compared with its 1st order moment of
quadratic distance $\langle\Delta X\rangle^{2}$ (squared mean), where
$\langle...\rangle$ represents the quantum expectation value by the
state of the reference system $|X\rangle$. In this 1st order approximation,
this quantum framework recovers the standard textbook quantum mechanics
without gravity. When the quantum fluctuation $\langle\delta X^{2}\rangle$
as the 2nd order correction of the reference frame system is important
and taken into account, gravity as a next order effects emerges in
the quantum framework, as if one introduces gravitation into the standard
textbook quantum mechanics, details are seen below and in previous
works.

To find the state $|X\rangle\in\mathcal{H}_{X}$ of the quantum reference
system, a quantum theory of the reference frame must be introduced.
If the quantum spacetime reference frame $|X^{\mu}\rangle$ ($\mu=0,1,2,...D-1$)
itself is considered as the to-be-studied quantum system, w.r.t. the
fiducial lab spacetime $|x_{a}\rangle$ as the reference system, ($a=0,1,2,...d-1$),
the entangled state $|X(x)\rangle=\sum_{ij}\alpha_{ij}|X\rangle_{i}\otimes|x\rangle_{j}$
can be constructed by a mapping between the two states, i.e. $|x\rangle\rightarrow|X\rangle$.
From the mathematical viewing point, to defined a D-dimensional manifolds
we need to construct a non-linear differentiable mapping $X(x)$ from
a local coordinate patch $x\in\mathbb{R}^{d}$ to a D-manifolds $X\in M^{D}$.
The mapping in physics is usually realized by a kind of fields theory
for $X(x)$, the non-linear sigma model (NLSM) \citep{gell1960axial,friedan1980nonlinear,Friedan1980,zinn2002quantum,codello2009fixed,percacci2009asymptotic,ketov2013quantum,2015Non}
\begin{equation}
S[X]=\frac{1}{2}\lambda\sum_{\mu,\nu=0}^{D-1}\int d^{d}xg_{\mu\nu}\sum_{a=0}^{d-1}\frac{\partial X^{\mu}}{\partial x_{a}}\frac{\partial X^{\nu}}{\partial x_{a}},\label{eq:NLSM}
\end{equation}
where $\lambda$ is a constant with dimension of energy density $[L^{-d}]$
taking the value of the critical density (\ref{eq:critical density})
of the universe. 

In the action, $x_{a}$ $(a=0,1,2,...,d-1)$, with dimension length
$[L]$, is called the base space in NLSM's terminology, representing
the coordinates of the local patch. They will be interpreted as the
lab wall and clock frame as the starting reference, which is considered
fiducial and classical with infinite precision. For the reason that
a quantum fields theory must be formulated in a classical inertial
frame, i.e. flat Minkowskian or Euclidean spacetime, so the base space
is considered flat. Without loss of generality, we consider the base
space as the Euclidean one, i.e. $x\in\mathbb{R}^{d}$ which is better
defined when one tries to quantize the theory. 

The differential mapping $X_{\mu}(x)$, $(\mu=0,1,2,...,D-1)$, with
dimensional length $[L]$, is the coordinates of a general Riemannian
or Lorentzian manifolds $M^{D}$ (depending on the boundary condition)
with curved metric $g_{\mu\nu}$, called the target space in NLSM's
terminology. We will work with the real-defined coordinates for the
target spacetime, and the Wick rotated version has been included into
the general coordinates transformation of the time component. In the
language of quantum fields theory, $X_{\mu}(x)$ or $X^{\mu}(x)=\sum_{\nu=0}^{D-1}g^{\mu\nu}X_{\nu}(x)$
are the real scalar frame fields.

Here, if not specifically mentioned, we will use the Einstein summation
convention to sum over index variable appears twice (Latin index for
the lab frame from 0 to $d-1$ and Greek index for the spacetime from
0 to $D-1$) and dropping the summation notation sigma. 

From the physical point of view, the reference frame fields can be
interpreted as a physical coordinates system by using particle/fields
signals, for instance, a multi-wire proportional chamber that measuring
coordinates of an event in a lab. To build a coordinates system, first
we need to orient, align and order the array of the multi-wires with
the reference to the wall of the lab $x_{a}$, ($a=1,2,3$). The electron
fields (ignoring the spin) in these array of multi-wires are considered
as the scalar frame fields. With the reference to the wall of the
lab, to locate a position of an event, at least three electron signals
$X_{1},X_{2},X_{3}$ must be received and read in three orthogonal
directions. The location information can be measured from the wave
function of the electron fields, e.g. from the phase counting or particle
number counting. Usually we could consider the electrons in the wires
are free, and the field's intensity is not very large, so that the
intensity can be seen as a linear function of the coordinates of the
lab's wall, $X_{\mu}(x)=\sum_{a=1}^{3}e_{\mu}^{a}x_{a}$, $(\mu=1,2,3)$,
for instance, here $e_{\mu}^{a}=\delta_{\mu}^{a}$ is the intensity
of the signals in each orthogonal direction. Meaning that when the
direction $\mu$ is the lab's wall direction $a$, the intensity of
the electron beam is 1, otherwise the intensity is 0. Similarly, one
need to read an extra electron signal $X_{0}$ to know when the event
happens, with the reference to the lab's clock $x_{0}$. Thus, the
fields of these 3+1 electron signals can be given by
\begin{equation}
X_{\mu}(x)=\sum_{a=0}^{3}e_{\mu}^{a}x_{a},\quad(\mu=0,1,2,3).\label{eq:X=00003Dex}
\end{equation}
The intensity of the fields $e_{\mu}^{a}$ is in fact the vierbein,
describing a mapping from the lab coordinate $x_{a}$ to the frame
fields $X_{\mu}$.

When the event happens at a long distance beyond the lab's scale,
for instance, at the scale of earth or solar system, we could imagine
that to extrapolate the multi-wire chamber to such long distance scale
still seems OK, only replacing the electrons beam in wire by the light
beam. However, if the scale is much larger than the solar system,
for instance, to the galaxy or cosmic scale, when the signal travels
along such long distance and be read by an observer, we could imagine
that the broadening of the light beam fields or other particle fields
gradually becomes non-negligible. More precisely, the 2nd (or higher)
order central moment fluctuations of the frame fields signals can
not be neglected, the distance of Riemannian/Lorentzian spacetime
as a quadratic form must be modified by the 2nd moment fluctuation
or variance $\langle\delta X^{2}\rangle$ of the coordinates
\begin{equation}
\left\langle \left(\varDelta X\right)^{2}\right\rangle =\langle\varDelta X\rangle^{2}+\langle\delta X^{2}\rangle.\label{eq:dx^2}
\end{equation}
A local distance element in spacetime is given by a local metric tensor
at the point, so it is convenient to think of the location point $X$
being fixed, and interpreting the variance of the coordinate affect
only the metric tensor $g_{\mu\nu}$ at the location point. As a consequence,
the expectation value of a metric tensor $g_{\mu\nu}$ is corrected
by the 2nd central moment quantum fluctuation of the frame fields
\begin{equation}
\langle g_{\mu\nu}\rangle=\left\langle \frac{\partial X_{\mu}}{\partial x_{a}}\frac{\partial X_{\nu}}{\partial x_{a}}\right\rangle =\left\langle \frac{\partial X_{\mu}}{\partial x_{a}}\right\rangle \left\langle \frac{\partial X_{\nu}}{\partial x_{a}}\right\rangle +\frac{1}{2}\frac{\partial^{2}}{\partial x_{a}^{2}}\left\langle \delta X_{\mu}\delta X_{\nu}\right\rangle =g_{\mu\nu}^{(1)}(X)-\delta g_{\mu\nu}^{(2)}(X),\label{eq:g=00003Dg(1)+dg(2)}
\end{equation}
where
\begin{equation}
g_{\mu\nu}^{(1)}(X)=\left\langle \frac{\partial X_{\mu}}{\partial x_{a}}\right\rangle \left\langle \frac{\partial X_{\nu}}{\partial x_{a}}\right\rangle =\langle e_{\mu}^{a}\rangle\langle e_{\nu}^{a}\rangle
\end{equation}
is the 1st order moment (mean value) contribution to the classical
spacetime. For the contribution of the 2nd order central moment $\delta g_{\mu\nu}^{(2)}$
(variance), the expectation value of the metric generally tends to
be curved up and deformed, the longer the distance scale the more
important the broadening of the frame fields, making the spacetime
geometry gradually deform and flow at long distance scale. 

Since the classical solution of the frame fields (\ref{eq:X=00003Dex})
given by the vierbein satisfying the classical equation of motion
of the NLSM, it is a frame fields interpretation of NLSM in a lab:
the base space of NLSM is interpreted as a starting reference by the
lab's wall and clock, the frame fields $X(x)$ on the lab are the
physical instruments measuring the spacetime coordinates. In this
interpretation we consider $d=4-\epsilon$, $(0<\epsilon\ll1)$ in
(\ref{eq:NLSM}) and $D=4$ is the least number of the frame fields.

There are several reason why $d$ is not precise but very close to
4 in the quantum frame fields interpretation of NLSM. $d$ must be
very close to 4, first, certainly at the scale of lab it is our common
sense; Second if we consider the entangled system $\mathcal{H}_{\psi}\otimes\mathcal{H}_{X}$
between the to-be-studied physical system and the reference frame
fields system, without loss of generality, we could take a scalar
field $\psi$ as the to-be-studied (matter) system, which shares the
common base space with the frame fields, the total action of the two
entangled system is a direct sum of each system
\begin{equation}
S[\psi,X]=\int d^{d}x\left[\frac{1}{2}\frac{\partial\psi}{\partial x_{a}}\frac{\partial\psi}{\partial x_{a}}-V(\psi)+\frac{1}{2}\lambda g_{\mu\nu}\frac{\partial X^{\mu}}{\partial x_{a}}\frac{\partial X^{\nu}}{\partial x_{a}}\right],
\end{equation}
where $V(\psi)$ is some potential of the $\psi$ fields. It can be
interpreted as an action of a quantum fields $\psi$ on general spacetime
coordinates $X$. Since both $\psi$ field and the frame fields $X$
share the same base space $x$, here they are described w.r.t. the
lab spacetime $x$ as the textbook quantum fields theory defined on
inertial frame $x$. If we interpret the frame fields as the physical
general spacetime coordinates, the coordinates of $\psi$ field must
be transformed from inertial frame $x$ to general coordinates $X$.
At the semi-classical level, or 1st moment approximation when the
fluctuation of $X$ can be ignored, it is simply a classical coordinates
transformation
\begin{align}
S[\psi,X]\overset{(1)}{\approx}S[\psi(X)] & =\int d^{4}X\sqrt{|\det g^{(1)}|}\left[\frac{1}{4}\left\langle g_{\mu\nu}^{(1)}\frac{\partial X^{\mu}}{\partial x_{a}}\frac{\partial X^{\nu}}{\partial x_{a}}\right\rangle \left(\frac{1}{2}g^{(1)\mu\nu}\frac{\delta\psi}{\delta X^{\mu}}\frac{\delta\psi}{\delta X^{\nu}}+2\lambda\right)-V(\psi)\right]\nonumber \\
 & =\int d^{4}X\sqrt{|\det g^{(1)}|}\left[\frac{1}{2}g^{(1)\mu\nu}\frac{\delta\psi}{\delta X^{\mu}}\frac{\delta\psi}{\delta X^{\nu}}-V(\psi)+2\lambda\right],\label{eq:coupled gravity (1)}
\end{align}
in which $\overset{(1)}{\approx}$ stands for the 1st moment or semi-classical
approximation, and $\frac{1}{4}\left\langle g_{\mu\nu}^{(1)}\frac{\partial X^{\mu}}{\partial x_{a}}\frac{\partial X^{\nu}}{\partial x_{a}}\right\rangle =\frac{1}{4}\left\langle g_{\mu\nu}^{(1)}g^{(1)\mu\nu}\right\rangle =\frac{1}{4}D=1$
has been used. It is easy to see, at the semi-classical level, i.e.
only consider the 1st moment of $X$ while 2nd moment fluctuations
are ignored, the (classical) coordinates transformation reproduces
the scalar field action in general coordinates $X$ up to a constant
$2\lambda$, and the derivative $\frac{\partial}{\partial x_{a}}$
is formally replaced by the functional derivative $\frac{\delta}{\delta X^{\mu}}$.
$\sqrt{|\det g^{(1)}|}$ is the Jacobian determinant of the coordinate
transformation, note that the determinant requires the coordinates
transformation matrix to be a square matrix, so at semi-classical
level $d$ must be very close to $D=4$, which is not necessarily
true beyond the semi-classical level, when the 2nd moment quantum
fluctuations are important. For instance, since $d$ is a parameter
but an observable in the theory, it could even not necessary be an
integer but effectively fractal at the quantum level. 

$d$ not precisely 4 is for the quantum and topological reasons. To
investigate this, we note that quantization depends on the homotopy
group $\pi_{d}(M^{D})$ of the mapping $X(x):\mathbb{R}^{d}\rightarrow M^{D}$.
If we consider the (Wick rotated) spacetime $M^{D}$ topologically
the $S^{D}$ for simplicity, the homotopy group is trivial for all
$d<D=4$, in other words, when $d<4$ the mapping $X(x)$ will be
free from any unphysical singularities for topological reason, in
this situation, the target spacetime is always mathematically well-defined.
However, the situation $d=4$ is a little subtle, since $\pi_{4}(S^{4})=\mathbb{Z}$
is non-trivial, the mapping might meet intrinsic topological obstacle
and become singular, i.e. a singular spacetime configuration. When
the quantum principle is taken into account, this situation can not
be avoided, and by its RG flow the spacetime is possibly deformed
into intrinsic singularities making the theory ill-defined at the
quantum level and non-renormalizable (RG flow not converge). So at
the quantum level, $d=4$ should be not precisely, we have to assume
$d=4-\epsilon$ when the quantum principle applies, while at the classical
or semi-classical level, considering $d=4$ has no serious problem.
The above argument is different from the conventional simple power
counting argument, which claims the NLSM is perturbative non-renormalizable
when $d>2$, but it is not necessarily the case, it is known that
numerical calculations also support $d=3$ and $d=4-\epsilon$ are
non-perturbative renormalizable and well-defined at the quantum level.

\subsection{Beyond the Semi-Classical Level: Gaussian Approximation}

Going beyond the semi-classical or 1st order moment approximation,
we need to quantize the theory at least at the next leading order.
If we consider the 2nd order central moment quantum fluctuation are
the most important next leading order contribution (compared with
higher order moment), we call it the Gaussian approximation or 2nd
order central moment approximation, while the higher order moment
are all called non-Gaussian fluctuations which might be important
near local singularities of the spacetime when local phase transition
happens, although the intrinsic global singularity can be avoided
by guaranteeing the global homotopy group trivial.

At the Gaussian approximation, $\delta g_{\mu\nu}^{(2)}$ can be given
by a perturbative one-loop calculation \citep{codello2009fixed,percacci2009asymptotic}
of the NLSM when it is relatively small compared with $g_{\mu\nu}^{(1)}$
\begin{equation}
\delta g_{\mu\nu}^{(2)}(X)=\frac{R_{\mu\nu}^{(1)}(X)}{32\pi^{2}\lambda}\delta k^{2},\label{eq:dg(2)}
\end{equation}
where $R_{\mu\nu}^{(1)}$ is the Ricci curvature given by 1st order
metric $g_{\mu\nu}^{(1)}$, $k^{2}$ is the cutoff energy scale of
the Fourier component of the frame fields. The validity of the perturbation
calculation $R^{(1)}\delta k^{2}\ll\lambda$ is the validity of the
Gaussian approximation, which can be seen as follows. It will be shown
in later section that $\lambda$ is nothing but the critical density
$\rho_{c}$ of the universe, $\lambda\sim O(H_{0}^{2}/G)$, $H_{0}$
the Hubble's constant, $G$ the Newton's constant. Thus for our concern
of pure gravity in which matter is ignored, the condition $R^{(1)}\delta k^{2}\ll\lambda$
is equivalent to $\delta k^{2}\ll1/G$ which is reliable except for
some local singularities are developed when the Gaussian approximation
is failed.

The equation (\ref{eq:dg(2)}) is nothing but a RG equation or known
as the Ricci flow equation (some reviews see e.g. \citep{chow2004ricci,chow2006hamilton,topping2006lectures})
\begin{equation}
\frac{\partial g_{\mu\nu}}{\partial t}=-2R_{\mu\nu},\label{eq:ricci flow}
\end{equation}
with flow parameter $\delta t=-\frac{1}{64\pi^{2}\lambda}\delta k^{2}$
having dimension of length squared $[L^{2}]$, which continuously
deform the spacetime metric driven by its Ricci curvature. 

For the Ricci curvature is non-linear for the metric, the Ricci flow
equation is a non-linear version of a heat equation for the metric,
and flow along $t$ introduces an averaging or coarse-graining process
to the intrinsic non-linear gravitational system which is highly non-trivial
\citep{carfora1995renormalization,piotrkowska1995averaging,carfora2008ricci,Zalaletdinov:2008ts,Paranjape:2009zu}.
In general, if the flow is free from local singularities there exists
long flow-time solution in $t\in(-\infty,0)$, which is often called
ancient solution in mathematical literature. This range of the t-parameter
corresponds to $k\in(0,\infty)$, that is from $t=-\infty$, i.e.
the short distance (high energy) UV scale $k=\infty$ forwardly to
$t=0$ i.e. the long distance (low energy) IR scale $k=0$. The metric
at certain scale $t$ is given by being averaged out the shorter distance
details which produces an effective correction to the metric at that
scale. So along t, the manifolds loss its information in shorter distance,
thus the flow is irreversible, i.e. generally having no backwards
solution, which is the underlying reason for the non-unitary and existence
of entropy of a spacetime. 

As it is shown in (\ref{eq:dx^2}), (\ref{eq:g=00003Dg(1)+dg(2)}),
the 2nd order moment fluctuation modifies the local (quadratic) distance
of the spacetime, so the flow is non-isometry. This is an important
feature worth stressing, which is the underlying reason for the anomaly.
The non-isometry is not important for its topology, so along t, the
flow preserves the topology of the spacetime but its local metric,
shape and size (volume) changes. There also exists a very special
solution of the Ricci flow called Ricci Soliton, which only changes
the local volume while keeps its local shape. The Ricci Soliton, and
its generalized version, the Gradient Ricci Soliton, as the flow limits,
are the generalization of the notion of fixed point in the sense of
RG flow. The Ricci Soliton is an important notion for understanding
the gravity at cosmic scale and studying the the thermodynamics of
the Ricci flow at equilibrium.

The Ricci flow was initially introduced in 1980s by Friedan \citep{friedan1980nonlinear,Friedan1980}
in $d=2+\epsilon$ NLSM and independently by Hamilton in mathematics
\citep{Hamilton1982Three,hamilton1986four}. The main motivation of
introducing it from the mathematical point of view is to classify
manifolds, a specific goals is to proof the Poincare conjecture. Hamilton
used it as a useful tool to gradually deform a manifolds into a more
and more ``simple and good'' manifolds whose topology can be readily
recognized for some simple cases. A general realization of the program
is achieved by Perelman at around 2003 \citep{perelman2002entropy,perelman2003ricci,perelman307245finite},
who introduced several monotonic functionals to successfully deal
with the local singularities which might be developed in more general
cases. The Ricci flow approach is not only powerful to the compact
geometry (as Hamilton's and Perelman's seminal works had shown) but
also to the non-compact \citep{1989Ricci,1989Deforming,2005Uniqueness}
and the Lorentzian geometry \citep{2010arXiv1007.3397B,2011arXiv1106.2924B,Ruchin:2013azz,2014arXiv1403.4400B,2019arXiv190703541B,2020AnPhy.42368333B,2020EPJC...80..639V,Luo:2022goc}.

\subsection{The Wavefunction and Density Matrix at the Gaussian Approximation }

So far we have not explicitly defined the quantum state of the reference
frame $|X\rangle$ in (\ref{eq:entangle}). In fact, the previous
(2nd order) results e.g. (\ref{eq:g=00003Dg(1)+dg(2)}), (\ref{eq:dg(2)})
and hence the Ricci flow (\ref{eq:ricci flow}) can also equivalently
be given by the expectation value $\langle O\rangle=\langle X|O|X\rangle$
via explicitly writing down the wavefunction $\Psi(X)$ of the frame
fields at the Gaussian approximation. Note that at the semi-classical
level, the frame fields $X$ is a delta-distribution and peaks at
its mean value, and further more, the action of the NLSM seems like
a collection of harmonic oscillators, thus at the Gaussian approximation
level, finite Gaussian width/2nd moment fluctuation of $X$ must be
introduced. When one performs a canonical quantization to the NLSM
at the Gaussian approximation level, the fundamental solution of the
wave function(al) (as a functional of the frame fields $X^{\mu}$)
of NLSM takes the Gaussian form, i.e. a coherent state
\begin{equation}
\Psi[X^{\mu}(x)]=\frac{1}{\sqrt{\lambda}(2\pi)^{D/4}}\frac{\left|\det\sigma_{\mu\nu}\right|^{1/4}}{|\det g_{\mu\nu}|^{1/4}}\exp\left[-\frac{1}{4}\left|X^{\mu}(x)\sigma_{\mu\nu}(x)X^{\nu}(x)\right|\right],
\end{equation}
where the covariant matrix $\sigma_{\mu\nu}(x)$, playing the role
of the Gaussian width, is the inverse of the 2nd order central moment
fluctuations of the frame fields at point $x$
\begin{equation}
\sigma_{\mu\nu}(x)=\frac{1}{\sigma^{\mu\nu}(x)}=\frac{1}{\left\langle \delta X^{\mu}(x)\delta X^{\nu}(x)\right\rangle },
\end{equation}
which is also given by perturbative one-loop calculation up to a diffeomorphism
of $X$. The absolute symbol of $\left|X^{\mu}\sigma_{\mu\nu}X^{\nu}\right|$
in the exponential is used to guarantee the quadratic form and hence
the determinant of $\sigma_{\mu\nu}$ induced from the Gaussian integral
over $X$ positive even in the Lorentzian signature.

We can also define a dimensionless density matrix corresponding to
the fundamental solution of the wavefunction
\begin{equation}
u[X^{\mu}(x)]=\Psi^{*}(X)\Psi(X)=\frac{1}{\lambda(2\pi)^{D/2}}\frac{\sqrt{\left|\det\sigma_{\mu\nu}\right|}}{\sqrt{|\det g_{\mu\nu}|}}\exp\left[-\frac{1}{2}\left|X^{\mu}(x)\sigma_{\mu\nu}X^{\nu}(x)\right|\right],\label{eq:u}
\end{equation}
and $\frac{1}{\lambda(2\pi)^{D/2}}\frac{\sqrt{\left|\det\sigma_{\mu\nu}\right|}}{\sqrt{|\det g_{\mu\nu}|}}$
is a normalization parameter, so that
\begin{equation}
\lambda\int d^{D}X\Psi^{*}(X)\Psi(X)=\lambda\int d^{D}Xu(X)=1,\label{eq:u-normalization}
\end{equation}
in which we often attribute the flow of the volume form $d^{D}X_{t}$
to the flow of the metric $g_{t}$, for the volume element $d^{D}X_{t}\equiv dV_{t}(X^{\mu})\equiv\sqrt{|g_{t}|}dX^{0}dX^{1}dX^{2}dX^{3}$.
Then the expectation values $\langle O\rangle$ can be understood
as $\lambda\int d^{D}X_{t}uO$. As the quantum frame fields $X$ are
q-number in the theory, precisely speaking, the integral of them should
be, in principle, a functional integral. Here the formal c-number
integral of them $\int d^{D}X_{t}...$ is for the conventional in
the Ricci flow literature, in which $X$ is a coarse-grained c-number
coordinates of manifolds at scale $t$. The exact functional integral
of $X$ is considered in calculating the partition function and related
anomaly of the theory in section-III.

Under a diffeomorphism of the metric, the transformation of $u(X)$
is given by a diffeomorphism of the covariant matrix ($h$ is certain
function)
\begin{equation}
\sigma_{\mu\nu}\rightarrow\hat{\sigma}_{\mu\nu}=\sigma_{\mu\nu}+\nabla_{\mu}\nabla_{\nu}h.
\end{equation}
So there exists an arbitrariness in the density $u(X)$ for different
choices of a diffeomorphism/gauge.

According to the statistical interpretation of wavefunction with the
normalization condition (\ref{eq:u-normalization}), $u(X^{0},X^{1},X^{2},X^{3})$
describes the probability density that finding these frame particles
in the volume $dV_{t}(X^{\mu})$. As the spacetime $X$ flows along
$t$, the volume $\Delta V_{t}$, in which density is averaged, also
flows, so the density at the corresponding scale is coarse-grained.
If we consider the volume of the lab, i.e the base space, is rigid
and fixed by $\lambda\int d^{4}x=1$, by noting (\ref{eq:u-normalization}),
we have
\begin{equation}
u[X^{\mu}(x),t]=\frac{d^{4}x}{d^{D}X_{t}}=\lim_{\Delta V_{t}\rightarrow0}\frac{1}{\Delta V_{t}}\int_{\Delta V_{t}}1\cdot d^{4}x.\label{eq:coarse-grain density}
\end{equation}
We can see that the density $u(X,t)$ can be interpreted as a coarse-grained
density in the volume element $\Delta V_{t}\rightarrow0$ w.r.t. a
fine-grained unit density in the lab volume element $d^{4}x$ at UV
$t\rightarrow-\infty$.

In this sense, the coarse-grained density $u(X,t)$ is in analogy
with the Boltzmann's distribution function, so it should satisfy an
analogous irreversible Boltzmann's equation, and giving rise to an
analogous Boltzmann's monotonic H-functional. In the following sections,
we will deduce such equation and the functional of $u(X,t)$. The
coarse-grained density $u(X,t)$ has profound physical and geometric
meaning, it also plays a central role in analyzing the statistic physics
of the frame fields and generalizes the manifolds to the density manifolds.

\subsection{Ricci-DeTurck Flow}

In previous subsection, from the viewpoint of frame fields particle,
$u(X^{\mu},t)$ has a coarse-grained particle density interpretation,
the eq.(\ref{eq:coarse-grain density}) can also be interpreted as
a manifolds density \citep{2016arXiv160208000W} from the geometric
point of view. For instance, $u(X,t)$ associates a manifold density
or density bundle to each point $X$ of a manifolds, measures the
fuzziness of the ``point''. It is worth stressing that the manifolds
density $u(X,t)$ does not simply a conformal scaling of a metric
by the factor, since if it is the case, the integral measure of $D=4$-volume
or 3-volume in the expectation $\langle O\rangle=\lambda\int d^{D}XuO$
would scale by different powers. There are various useful generalizations
of the Ricci curvature to the density manifolds, a widely accepted
version is the Bakry-Emery generalization \citep{1985Diffusions}
\begin{equation}
R_{\mu\nu}\rightarrow R_{\mu\nu}-\nabla_{\mu}\nabla_{\nu}\log u,
\end{equation}
which is also used in Perelman's seminal paper. The density normalized
Ricci curvature is bounded from below
\begin{equation}
R_{\mu\nu}-\nabla_{\mu}\nabla_{\nu}\log u\ge\sigma_{\mu\nu},\label{eq:density ric bound}
\end{equation}
if the density manifolds has finite volume.

As a consequence, replacing the Ricci curvature by the density normalized
one, we get the Ricci-DeTurck flow \citep{deturck1983deforming}
\begin{equation}
\frac{\partial g_{\mu\nu}}{\partial t}=-2\left(R_{\mu\nu}-\nabla_{\mu}\nabla_{\nu}\log u\right),\label{eq:ricci-deturk}
\end{equation}
which is equivalent to the standard Ricci flow equation (\ref{eq:ricci flow})
up to a diffeomorphism. Mathematically, the Ricci-DeTurck flow has
the advantage that it turns out to be a gradient flow of some monotonic
functionals introduced by Perelman, which have profound physical meanings
shown later.

The eq.(\ref{eq:u-normalization}) and (\ref{eq:coarse-grain density})
also give a volume constraint to the fiducial spacetime (the lab),
the coarse-grained density $u(X,t)$ cancels the flow of the volume
element $\sqrt{|\det g_{\mu\nu}|}$, so
\begin{align}
\frac{\partial}{\partial t}\left(u\sqrt{|\det g_{\mu\nu}|}\right) & =0.\label{eq:volume constraint}
\end{align}
Together with the Ricci-DeTurck flow equation (\ref{eq:ricci-deturk}),
we have the flow equation of the density
\begin{equation}
\frac{\partial u}{\partial t}=\left(R-\Delta_{X}\right)u,
\end{equation}
which is in analogy to the irreversible Boltzmann's equation for his
distribution function. $\Delta_{X}$ is the Laplacian operator in
terms of the manifolds coordinates $X$. Note the minus sign in front
of the Laplacian, it is a backwards heat-like equation. Naively speaking,
the solution of the backwards heat flow will not exist. But we could
also note that if one let the Ricci flow flows to certain IR scale
$t_{*}$, and at $t_{*}$ one might then choose an appropriate $u(t_{*})=u_{0}$
arbitrarily (up to a diffeomorphism gauge) and flows it backwards
in $\tau=t_{*}-t$ to obtain a solution $u(\tau)$ of the backwards
equation. Now since the flow is consider free from global singularities
for the trivialness of the homotopy group, we could simply choose
$t_{*}=0$, so we defined
\begin{equation}
\tau=-t=\frac{1}{64\pi^{2}\lambda}k^{2}\in(0,\infty).\label{eq:tau}
\end{equation}
In this case, the density satisfies the heat-like equation
\begin{equation}
\frac{\partial u}{\partial\tau}=\left(\Delta_{X}-R\right)u,\label{eq:u-equation}
\end{equation}
which does admit a solution along $\tau$, often called the conjugate
heat equation in mathematical literature. 

So far (\ref{eq:u-equation}) together with (\ref{eq:ricci-deturk})
the mathematical problem of the Ricci flow of a Riemannian/Lorentzian
manifolds is transformed to a coupled equations
\begin{equation}
\begin{cases}
\frac{\partial g_{\mu\nu}}{\partial t}=-2\left(R_{\mu\nu}-\nabla_{\mu}\nabla_{\nu}\log u\right)\\
\frac{\partial u}{\partial\tau}=\left(\Delta_{X}-R\right)u\\
\frac{d\tau}{dt}=-1
\end{cases}
\end{equation}
and the manifolds $(M^{D},g)$ is generalized to a density manifolds
$(M^{D},g,u)$ \citep{Morgan2009Manifolds,2016arXiv160208000W,Corwin2017Differential}
with the constraint (\ref{eq:u-normalization}).

\section{The Anomaly and its Implications}

At the semi-classical approximation, see in eq.(\ref{eq:coupled gravity (1)}),
when the quantum fluctuations of the frame fields or spacetime coordinates
are ignored, the general coordinates transformation is just a classical
coordinates transformation. We will show that when the quantum fluctuations
are taken into account in the general coordinates transformation beyond
the semi-classical approximation, quantum anomaly emerges. As is seen
in the previous section, the quantum fluctuation and hence the coarse-graining
process of the Ricci flow does not preserve the quadratic distance
of a geometry, see (\ref{eq:dx^2}) and (\ref{eq:g=00003Dg(1)+dg(2)}).
The non-isometry of the quantum fluctuation induces a breakdown of
diffeomorphism or general coordinate transformation at the quantum
level, namely the diffeomorphism anomaly. In this section, we derive
the diffeomorphism anomaly of the theory, show its relation to the
Shannon entropy whose monotonicity gives an analogous H-theorem of
the frame fields system and the Ricci flow. Further more, as the quantum
frame fields theory describes a quantum spacetime, together with the
generalized quantum Equivalence Principle, the anomaly induced effective
action in terms of the Shannon entropy can also be interpreted as
a gravity theory, which at low energy expansion is a scale dependent
Einstein-Hilbert action plus a cosmological constant. This part has
certain overlap with the previous work \citep{Luo:2021zpi}, for the
self-containedness of the paper, we hope this section provide a general
background and lay the foundation for the subsequent thermodynamic
and statistic interpretation of the theory.

\subsection{Diffeomorphism at the Quantum Level}

First we consider the functional quantization of the pure frame fields
without explicitly incorporating the matter source. The partition
function is
\begin{equation}
Z(M^{D})=\int[\mathcal{D}X]\exp\left(-S[X]\right)=\int[\mathcal{D}X]\exp\left(-\frac{1}{2}\lambda\int d^{4}xg^{\mu\nu}\partial_{a}X_{\mu}\partial_{a}X_{\nu}\right),\label{eq:partition of NLSM}
\end{equation}
where $M^{D}$ is the target spacetime, and the base space can be
either Euclidean and Minkowskian. Since considering the action or
the volume element $d^{4}x\equiv d^{4}x\det e$ ($\det e$ is a Jacobian)
does not pick any imaginary $i$ factor no matter the base space is
in Minkowskian or Euclidean one, if one takes $dx_{0}^{(E)}\rightarrow idx_{0}^{(M)}$
then $\det e^{(E)}\rightarrow-i\det e^{(M)}$, so without loss of
generality we use the Euclidean base spacetime in the following discussions,
and remind that the result is the same for Minkowskian. 

Note that a general coordinate transformation
\begin{equation}
X_{\mu}\rightarrow\hat{X}_{\mu}=\frac{\partial\hat{X}_{\mu}}{\partial X_{\nu}}X_{\nu}=e_{\mu}^{\nu}X_{\nu}
\end{equation}
does not change the action $S[X]=S[\hat{X}]$, but the measure of
the functional integral changes
\begin{align}
\mathcal{D}\hat{X} & =\prod_{x}\prod_{\mu=0}^{D-1}d\hat{X}_{\mu}(x)=\prod_{x}\epsilon_{\mu\nu\rho\sigma}e_{\mu}^{0}e_{\nu}^{1}e_{\rho}^{2}e_{\sigma}^{3}dX_{0}(x)dX_{1}(x)dX_{2}(x)dX_{3}(x)\nonumber \\
 & =\prod_{x}\left|\det e(x)\right|\prod_{x}\prod_{a=0}^{D-1}dX_{a}(x)=\left(\prod_{x}\left|\det e(x)\right|\right)\mathcal{D}X,
\end{align}
where
\begin{equation}
\epsilon_{\mu\nu\rho\sigma}e_{\mu}^{0}e_{\nu}^{1}e_{\rho}^{2}e_{\sigma}^{3}=\left|\det e_{\mu}^{a}\right|=\sqrt{\left|\det g_{\mu\nu}\right|}
\end{equation}
is the Jacobian of the diffeomorphism. The Jacobian is nothing but
a local relative (covariant basis) volume element $dV(\hat{X}_{\mu})$
w.r.t. the fiducial volume $dV(X_{a})$. Note that the normalization
condition (\ref{eq:u-normalization}) also defines a fiducial volume
element $ud^{4}X\equiv udV(\hat{X}_{\mu})$, so the Jacobian is nothing
but related to the frame fields density matrix
\begin{equation}
u(\hat{X}_{\mu})=\frac{dV(X_{a})}{dV(\hat{X}_{\mu})}=\left|\det e_{a}^{\mu}\right|=\frac{1}{\left|\det e_{\mu}^{a}\right|}.\label{eq:volume form}
\end{equation}

Here the absolute symbol of the determinant is because the density
$u$ and the volume element are kept positive defined even in the
Lorentz signature. Otherwise, for the Lorentz signature, it should
introduce some extra imaginary factor $i$ into (\ref{eq:parametrize u})
to keep the condition (\ref{eq:u-normalization}). The density so
defined followed by (\ref{eq:u-normalization}) is an explicit generalization
from the standard 3-space density to a 4-spacetime version. It is
the definition of the volume form and the manifolds density ensure
the formalism of the framework formally the same with the Perelman's
standard form even in the Lorentzian signature. The manifolds density
encodes the most important information of a Riemannian or Lorentzian
geometry, i.e. the local volume comparison.

In this case, if we parameterize a dimensionless solution $u$ of
the conjugate heat equation as
\begin{equation}
u(\hat{X})=\frac{1}{\lambda(4\pi\tau)^{D/2}}e^{-f(\hat{X})},\label{eq:parametrize u}
\end{equation}
then the partition function $Z(M^{D})$ is transformed to
\begin{align}
Z(\hat{M}^{D}) & =\int[\mathcal{D}\hat{X}]\exp\left(-S[\hat{X}]\right)=\int\left(\prod_{x}\left|\det e\right|\right)[\mathcal{D}X]\exp\left(-S[X]\right)\nonumber \\
 & =\int\left(\prod_{x}e^{f+\frac{D}{2}\log(4\pi\tau)}\right)[\mathcal{D}X]\exp\left(-S[X]\right)\nonumber \\
 & =\exp\left(\lambda\int d^{4}x\left[f+\frac{D}{2}\log(4\pi\tau)\right]\right)\int[\mathcal{D}X]\exp\left(-S[X]\right)\nonumber \\
 & =\exp\left(\lambda\int_{\hat{M}^{D}}d^{D}Xu\left[f+\frac{D}{2}\log(4\pi\tau)\right]\right)\int[\mathcal{D}X]\exp\left(-S[X]\right).
\end{align}

Note that $N(\hat{M}^{D})$ in the exponential of the change of the
partition function 
\begin{equation}
Z(\hat{M}^{D})=e^{\lambda N(\hat{M}^{D})}Z(M^{D})\label{eq:Z->Zhat}
\end{equation}
is nothing but a pure real Shannon entropy in terms of the density
matrix $u$
\begin{equation}
N(\hat{M}^{D})=\int_{\hat{M}^{D}}d^{D}Xu\left[f+\frac{D}{2}\log(4\pi\tau)\right]=-\int_{\hat{M}^{D}}d^{D}Xu\log u.
\end{equation}

The classical action $S[X]$ is invariant under the general coordinates
transformation or diffeomorphism, but the quantum partition function
is no longer invariant under the general coordinates transformation
or diffeomorphism, which is called diffeomorphism anomaly, meaning
a breaking down of the diffeomorphism at the quantum level. The diffeomorphism
anomaly is purely due to the quantum fluctuation and Ricci flow of
the frame fields which do not preserve the functional integral measure
and change the spacetime volume at the quantum level. The diffeomorphism
anomaly has many profound consequences to the theory of quantum reference
frame, e.g. non-unitarity, the trace anomaly, the notion of entropy,
reversibility, and the cosmological constant. 

The non-unitarity is indicated by the pure real anomaly term, which
is also induced by the non-isometry or volume change, and consequently
the non-invariance of the measure of the functional integral during
the Ricci flow. Because of the real-defined volume form (\ref{eq:volume form})
for both Euclidean and Lorentzian signature, the pure real contribution
of the anomaly and hence the non-unitarity are valid not only for
spacetime with Euclidean but also for the Lorentzian signature, it
is a rather general consequence of the Ricci flow of spacetime. Essentially
speaking, the reason of the non-unitarity is because we have enlarged
the Hilbert space of the reference frame, from a rigid classical frame
to a fluctuating quantum frame. The non-unitarity implies the breakdown
of the fundamental Schrodinger equation which is only valid on a classical
time of inertial frame, the solution of which is in $\mathcal{H}_{\psi}$.
A fundamental equation playing the role of the Schrodinger equation,
which can arbitrarily choose any (quantum) physical system as time
or reference frame, must be replaced by a Wheeler-DeWitt-like equation
in certain sense \citep{Luo2014The}, the solution of which is instead
in $\mathcal{H}_{\psi}\otimes\mathcal{H}_{X}$. In the fundamental
equation, the quantum fluctuation of physical time and frame, more
generally, a general physical coordinates system must break the unitarity.
We know that in quantum fields theory on curved spacetime or accelerating
frame, the vacuum states of the quantum fields in difference diffeomorphism
equivalent coordinate systems are unitarily inequivalent. The Unruh
effect is a well known example: accelerating observers in the vacuum
will measure a thermal bath of particles. The Unruh effect shows us
how a general coordinates transformation (e.g. from an inertial to
an accelerating frame) leads to the non-unitary anomaly (particle
creation and hence particle number non-conservation), and how the
anomaly will relate to a thermodynamics system (thermal bath). In
fact, like the Unruh effect, the Hawking effect \citep{1977Trace}
and all non-unitary particle creation effects in a curved spacetime
or accelerating frame are related to the anomaly in a general covariant
or gravitational system. All these imply that the diffeomorphism anomaly
will have deep thermodynamic interpretation which is the central issue
of the paper. 

Without loss of generality, if we simply consider the under-transformed
coordinates $X_{\mu}$ identifying with the coordinates of the fiducial
lab $x_{a}$ which can be treated as a classical parameter coordinates,
in this situation the classical action of NLSM is just a topological
invariant, i.e. half the dimension of the target spacetime
\begin{equation}
\exp\left(-S_{cl}\right)=\exp\left(-\frac{1}{2}\lambda\int d^{4}xg^{\mu\nu}\partial_{a}x_{\mu}\partial_{a}x_{\nu}\right)=\exp\left(-\frac{1}{2}\lambda\int d^{4}xg^{\mu\nu}g_{\mu\nu}\right)=e^{-\frac{D}{2}}.
\end{equation}
Thus the total partition function of the frame fields takes a simple
form
\begin{equation}
Z(\hat{M}^{D})=e^{\lambda N(\hat{M}^{D})-\frac{D}{2}}.\label{eq:frame-partition}
\end{equation}

\subsection{The Trace Anomaly}

The partition function now is non-invariance (\ref{eq:Z->Zhat}) under
diffeomorphism at the quantum level, so if one deduces the stress
tensor by $\langle\mathcal{T}_{\mu\nu}\rangle=-\frac{2}{\sqrt{|g|}}\frac{\delta\log Z}{\delta g^{\mu\nu}}$,
its trace $\langle g^{\mu\nu}\rangle\langle\mathcal{T}_{\mu\nu}\rangle=0$
is difference from $\langle\mathcal{T}_{\mu}^{\mu}\rangle=\langle g^{\mu\nu}\mathcal{T}_{\mu\nu}\rangle$
\begin{equation}
\langle\Delta\mathcal{T}_{\mu}^{\mu}\rangle=\langle g^{\mu\nu}\rangle\langle\mathcal{T}_{\mu\nu}\rangle-\langle g^{\mu\nu}\mathcal{T}_{\mu\nu}\rangle=\lambda N(M^{D})\label{eq:trace anomaly}
\end{equation}
known as the trace anomaly. Cardy conjectured \citep{1988Is} that
in a $d=4$ theory, quantities like $\langle\mathcal{T}_{\mu}^{\mu}\rangle$
could be a higher dimensional generalization of the monotonic Zamolodchikov's
c-function in $d=2$ conformal theories, leading to a suggestion of
the a-theorem \citep{2011On} in $d=4$ and other suggestions (e.g.
\citep{1993Quantum,Gaite:1995yg}). In the following subsections,
we will show that the Shannon entropy $N$ and generalized $\tilde{N}$
are indeed monotonic, which might have more advantages, e.g. suitable
for a Lorentzian target spacetime and for general $D$.

Note that the Shannon entropy $N(M^{D})$ can be expanded at small
$\tau$ 
\begin{equation}
\lambda N(\hat{M}^{D})=\lambda\sum_{n=0}^{\infty}B_{n}\tau^{n}=\lambda\left(B_{0}+B_{1}\tau+B_{2}\tau^{2}+...\right)\quad(\tau\rightarrow0).
\end{equation}
For $D=4$ the first few coefficients are
\begin{equation}
B_{0}=\lim_{\tau\rightarrow0}N=\frac{D}{2\lambda}\left[1+\log\left(\sqrt{\lambda}4\pi\tau\right)\right],
\end{equation}
\begin{equation}
B_{1}=\lim_{\tau\rightarrow0}\frac{dN}{d\tau}=\int_{\hat{M}^{4}}d^{4}X\sqrt{|g|}\left(R+\frac{D}{2\tau}\right),
\end{equation}
\begin{equation}
B_{2}=\lim_{\tau\rightarrow0}\frac{1}{2}\frac{d^{2}N}{d^{2}\tau}=-\int_{\hat{M}^{4}}d^{4}X\sqrt{|g|}\left|R_{\mu\nu}+\frac{1}{2\tau}g_{\mu\nu}\right|^{2},
\end{equation}
in which $B_{0}$ can be renormalized out, and a renormalized $B_{1}$
will contribute to the effective Einstein-Hilbert action of gravity,
see following subsection D. And $B_{2}$, as a portion of the full
anomaly, plays the role of the conformal/Weyl anomaly up to some total
divergence terms, for instance, $\varDelta R$ terms and the Gauss-Bonnet
invariant. That is, a non-vanishing $B_{2}$ term measures the broken
down of the conformal invariance of $M^{D=4}$, otherwise, a vanishing
$B_{2}$ means that the manifold is a gradient steady Ricci soliton
as the fixed point of the Ricci-DeTurck flow, which preserves it shape
(conformal invariant) during the flows.

We note that $B_{2}$ as the only dimensionless coefficient measures
the anomalous conformal modes, in this sense, $N(M^{D})$ indeed relates
to certain entropy. However, since the conformal transformation is
just a special coordinates transformation, thus it is clearly that
the single $B_{2}$ coefficient does not measure the total (general
coordinates transformation) anomalous modes. Obviously this theory
at $2<d=4-\epsilon$ is not conformal invariant, thus as the theory
flows along $t$, the degrees of freedom are gradually coarse-grained
and hence the modes-counting should also change with the flow and
the scale, as a consequence all coefficients $B_{n}$ in the series
and hence the total partition function $e^{\lambda N(M^{D})}$ should
measure the total anomalous modes at certain scale $\tau$, leading
to the full entropy and anomaly. 

Different from some classically conformal invariant theories, e.g.
the string theory, in which we only need to cancel a single scale-independent
$B_{k}$ coefficient in order to avoid conformal anomaly. As the theory
at higher than 2-dimension is not conformal invariant, the full scale-dependent
anomaly $N(M^{D})$ is required to be canceled at certain scale. Fortunately,
it will show in later subsection that a normalized full anomaly $\lambda\tilde{N}(M^{D})$
can converge at UV for its monotonicity, thus giving rise to a finite
counter term of order $O(\lambda)$ playing the role of a correct
cosmological constant. The idea that the trace anomaly might have
a relation to the cosmological constant is a recurring subject in
literature \citep{Bilic:2007gr,Tomboulis:1988gw,Antoniadis:1991fa,Antoniadis:1992hz,Salehi:2000eu},
in the framework, the cosmological constant is naturally emerged in
this way as the counter term of the trace anomaly (see subsection-D
or \citep{Luo:2021zpi}).

\subsection{Relative Shannon Entropy and a H-Theorem for Non-Equilibrium Frame
Fields }

In the Ricci flow limit, i.e. the Gradient Shrinking Ricci Soliton
(GSRS) configuration, the Shannon entropy $N$ taking its maximum
value $N_{*}$, it is similar with the thermodynamics system being
in a thermal equilibrium state where its entropy is also maximal.
In mathematical literature of Ricci flow, it is often defined a series
of relative formulae w.r.t. the extreme values taking by the flow
limit GSRS or analogous thermal equilibrium state denoted by a subscript
{*}. 

In GSRS, the covariance matrix $\sigma^{\mu\nu}$ as 2nd central moment
of the frame fields with a IR cutoff $k$ is simply proportional to
the metric
\begin{equation}
\frac{1}{2}\sigma_{*}^{\mu\nu}=\frac{1}{2}\langle\delta X^{\mu}\delta X^{\nu}\rangle=\frac{1}{2\lambda}g^{\mu\nu}\int_{0}^{|p|=k}\frac{d^{4}p}{(2\pi)^{4}}\frac{1}{p^{2}}=\frac{k^{2}}{64\pi^{2}\lambda}g^{\mu\nu}=\tau g^{\mu\nu},\label{eq:sigma=00003Dtau*g}
\end{equation}
and then
\begin{equation}
\sigma_{*\mu\nu}=\left(\sigma_{*}^{\mu\nu}\right)^{-1}=\frac{1}{2\tau}g_{\mu\nu},
\end{equation}
which means a uniform Gaussian broadening is achieved. And in this
gauge, only longitudinal part of fluctuation exists. 

When the density normalized Ricci curvature is completely given by
the longitudinal fluctuation $\sigma_{\mu\nu}$, i.e. the inequality
(\ref{eq:density ric bound}) saturates, giving a Gradient Shrinking
Ricci Soliton (GSRS) equation
\begin{equation}
R_{\mu\nu}+\nabla_{\mu}\nabla_{\nu}f=\frac{1}{2\tau}g_{\mu\nu}.\label{eq:shrinker}
\end{equation}
It means, on the one hand, for a general $f(X)=\frac{1}{2}\left|\sigma_{\mu\nu}X^{\mu}X^{\nu}\right|$,
so $R_{\mu\nu}$ seems vanish, so the standard Ricci flow equation
(\ref{eq:ricci flow}) terminates; and on the other hand, the Ricci-DeTurck
flow (\ref{eq:ricci-deturk}) only changes the longitudinal size or
volume of the manifolds but its shape keep unchanged, thus the GSRS
can also be seen stop changing, up to a size or volume rescaling.
Thus the GSRS is a flow limit and can be viewed as a generalized RG
fixed point.

In the following, we consider relative quantities w.r.t. the GSRS
configuration. Considering a general Gaussian density matrix
\begin{equation}
u(X)=\frac{1}{\lambda(2\pi)^{D/2}}\frac{\sqrt{\left|\det\sigma_{\mu\nu}\right|}}{\sqrt{|\det g_{\mu\nu}|}}\exp\left(-\frac{1}{2}\left|X^{\mu}\sigma_{\mu\nu}X^{\nu}\right|\right),\label{eq:general u}
\end{equation}
in GSRS limit it becomes 
\begin{equation}
u_{*}(X)=\frac{1}{\lambda(4\pi\tau)^{D/2}}\exp\left(-\frac{1}{4\tau}\left|X\right|^{2}\right).\label{eq:u*}
\end{equation}
Therefore, in GSRS, a relative density can be defined by the general
Gaussian density $u(X)$ relative to the density $u_{*}(X)$ in GSRS
\begin{equation}
\tilde{u}(X)=\frac{u}{u_{*}}.
\end{equation}
By using the relative density, a relative Shannon entropy $\tilde{N}$
can be defined by
\begin{equation}
\tilde{N}(M^{D})=-\int d^{D}X\tilde{u}\log\tilde{u}=-\int d^{D}Xu\log u+\int d^{D}Xu_{*}\log u_{*}=N-N_{*}=-\log Z_{P}\le0,\label{eq:perelman-partition}
\end{equation}
where $Z_{P}$ is nothing but the Perelman's partition function 
\begin{equation}
\log Z_{P}=\int_{M^{D}}d^{D}Xu\left(\frac{D}{2}-f\right)\ge0,
\end{equation}
and $N_{*}$ is the maximum Shannon entropy
\begin{equation}
N_{*}=-\int d^{D}Xu_{*}\log u_{*}=\int d^{D}Xu_{*}\frac{D}{2}\left[1+\log(\sqrt{\lambda}4\pi\tau)\right]=\frac{D}{2\lambda}\left[1+\log(\sqrt{\lambda}4\pi\tau)\right].
\end{equation}

Since the relative Shannon entropy and the anomaly term is pure real,
so the change of the partition function under diffeomorphism is non-unitary.
For the coarse-graining nature of the density $u$, it is proved that
the relative Shannon entropy is monotonic non-decreasing along the
Ricci flow (along $t$),
\begin{equation}
\frac{d\tilde{N}(\hat{M}^{D})}{dt}=-\tilde{\mathcal{F}}\ge0,\label{eq:analog H-theorem}
\end{equation}
where $\tilde{\mathcal{F}}=\mathcal{F}-\mathcal{F}_{*}\le0$ is the
GSRS-normalized F-functional of Perelman
\begin{equation}
\mathcal{F}=\frac{dN}{d\tau}=\int_{M^{D}}d^{D}Xu\left(R+\left|\nabla f\right|^{2}\right)
\end{equation}
with the maximum value (at GSRS limit)
\begin{equation}
\mathcal{F}_{*}\equiv\mathcal{F}(u_{*})=\frac{dN_{*}}{d\tau}=\frac{D}{2\lambda\tau}.\label{eq:F*}
\end{equation}

The inequality (\ref{eq:analog H-theorem}) gives an analogous H-theorem
to the non-equilibrium frame fields and the irreversible Ricci flow.
The entropy is non-decreasing along the Ricci flow making the flow
irreversible in many aspects similar with the processes of irreversible
thermodynamics, meaning that as the observation scale of the spacetime
flows from short to long distance scale, the process losses information
and the Shannon entropy increases. The equal sign in (\ref{eq:analog H-theorem})
can be taken when the spacetime configuration has flowed to a limit
known as a Gradient Shrinking Ricci Soliton (GSRS), when the Shannon
entropy takes its maximum value. Similarly, at the flow limit the
density matrix $u_{*}$ eq.(\ref{eq:u*}) takes the analogous standard
Maxwell-Boltzmann distribution.

\subsection{Effective Gravity at Cosmic Scale and the Cosmological Constant}

In terms of the relative Shannon entropy, the total partition function
(\ref{eq:frame-partition}) of the frame fields is normalized by the
GSRS extreme value 
\begin{equation}
Z(M^{D})=\frac{e^{\lambda N-\frac{D}{2}}}{e^{\lambda N_{*}}}=e^{\lambda\tilde{N}-\frac{D}{2}}=Z_{P}^{-\lambda}e^{-\frac{D}{2}}=\exp\left[\lambda\int_{M^{D}}d^{D}Xu\left(f-D\right)\right].\label{eq:relative-partition}
\end{equation}

The relative Shannon entropy $\tilde{N}$ as the anomaly vanishes
at GSRS or IR scale, however, it is non-zero at ordinary lab scale
up to UV where the fiducial volume of the lab is considered fixed
$\lambda\int d^{4}x=1$. The cancellation of the anomaly at the lab
scale up to UV is physically required, which leads to the counter
term $\nu(M_{\tau=\infty}^{D})$ or cosmological constant. The monotonicity
of $\tilde{N}$ eq.(\ref{eq:analog H-theorem}) and the W-functional
implies \citep{perelman2002entropy,2004The}
\begin{equation}
\nu(M_{\tau=\infty}^{D})=\lim_{\tau\rightarrow\infty}\lambda\tilde{N}(M^{D},u,\tau)=\lim_{\tau\rightarrow\infty}\lambda\mathcal{W}(M^{D},u,\tau)=\inf_{\tau}\lambda\mathcal{W}(M^{D},u,\tau)<0,
\end{equation}
where $\mathcal{W}$, the Perelman's W-functional, is the Legendre
transformation of $\tilde{N}$ w.r.t. $\tau^{-1}$,
\begin{equation}
\mathcal{W}\equiv\tau\frac{\partial\tilde{N}}{\partial\tau}+\tilde{N}=\tau\tilde{\mathcal{F}}+\tilde{N}=\frac{d}{d\tau}\left(\tau\tilde{N}\right).\label{eq:W=00003DtauF+N}
\end{equation}
In other words, the difference between the effective actions (relative
Shannon entropies) at UV and IR is finite
\begin{equation}
\nu=\lambda(\tilde{N}_{UV}-\tilde{N}_{IR})<0.
\end{equation}

Perelman used his analogies: the temperature $T\sim\tau$, the (relative)
internal energy $U\sim-\tau^{2}\tilde{\mathcal{F}}$, the thermodynamics
entropy $S\sim-\mathcal{W}$, and the free energy $F\sim\tau\tilde{N}$,
up to proportional balancing the dimensions on both sides of $\sim$,
the equation (\ref{eq:W=00003DtauF+N}) is in analogy to the thermodynamics
equation $U-TS=F$. So in this sense the W-functional is also called
the W-entropy. Whether the thermodynamic analogies are real and physical,
or just pure coincidences, is an important issue discussed in the
next sections.

In fact $e^{\nu}<1$ (usually called the Gaussian density \citep{cao2004gaussian,cao2009recent})
is a relative volume or the reduced volume $\tilde{V}(M_{\tau=\infty}^{D})$
of the backwards limit manifolds introduced by Perelman, or the inverse
of the initial condition of the manifolds density $u_{\tau=0}^{-1}$.
A finite value of it makes an initial spacetime with unit volume from
UV flow and converge to a finite $u_{\tau=0}$, and hence the manifolds
finally converges to a finite relative volume/reduced volume instead
of shrinking to a singular point at $\tau=0$.

As an example, for a homogeneous and isotropic universe for which
the sizes of space and time (with a ``ball'' radius $a_{\tau}$)
are on an equal footing, i.e. a late epoch FRW-like metric $ds^{2}=a_{\tau}^{2}(-dx_{0}^{2}+dx_{1}^{2}+dx_{2}^{2}+dx_{3}^{2})$,
which is a Lorentzian shrinking soliton configuration. Note that the
shrinking soliton equation $R_{\mu\nu}=\frac{1}{2\tau}g_{\mu\nu}$
it satisfies and its volume form (\ref{eq:volume form}) are independent
to the signature, so it can be approximately given by a 4-ball value
$\nu(B_{\infty}^{4})\approx-0.8$ \citep{Luo:2019iby,Luo:2021zpi}.

So the partition function, which is anomaly canceled at UV and having
a fixed-volume fiducial lab, is
\begin{equation}
Z(M^{D})=e^{\lambda\tilde{N}-\frac{D}{2}-\nu}.
\end{equation}
Since $\lim_{\tau\rightarrow0}\tilde{N}(M^{D})=0$, so at small $\tau$,
$\tilde{N}(M^{D})$ can be expanded by powers of $\tau$
\begin{align}
\tilde{N}(M^{D}) & =\frac{\partial\tilde{N}}{\partial\tau}\tau+O(\tau^{2})=\tau\tilde{\mathcal{F}}+O(\tau^{2})\nonumber \\
 & =\int_{M^{D}}d^{D}Xu_{\tau\rightarrow0}\left[\left(R_{\tau\rightarrow0}+\left|\nabla f_{\tau\rightarrow0}\right|^{2}-\frac{D}{2\tau}\right)\tau\right]+O(\tau^{2})\nonumber \\
 & =\int_{M^{D}}d^{D}Xu_{0}R_{0}\tau+O(\tau^{2}),
\end{align}
in which $\lambda\int d^{D}Xu_{\tau\rightarrow0}\tau\left|\nabla f_{\tau\rightarrow0}\right|^{2}=\frac{D}{2}$
(at GSRS) has been used.

For $D=4$ and small $\tau$, the effective action of $Z(M^{4})$
can be given by
\begin{equation}
-\log Z(M^{4})=S_{eff}\approx\int_{M^{4}}d^{4}Xu_{0}\left(2\lambda-\lambda R_{0}\tau+\lambda\nu\right)\quad(\textrm{small\,\ensuremath{\tau}}).
\end{equation}
Considering $u_{0}d^{4}X=\sqrt{|g_{t}|}dV=\sqrt{|g_{t}|}dX^{0}dX^{1}dX^{2}dX^{3}$
is the invariant volume element, and using (\ref{eq:tau}) to replace
$t$ or $\tau$ by cutoff scale $k$, we have
\begin{equation}
S_{eff}=\int_{M^{4}}dV\sqrt{|g_{k}|}\left(2\lambda-\frac{R_{0}}{64\pi^{2}}k^{2}+\lambda\nu\right)\quad(\textrm{small\,k}).\label{eq:eff-EH+cc}
\end{equation}

The effective action can be interpreted as a low energy effective
action of pure gravity. As the cutoff scale $k$ ranges from the lab
scale to the solar system scale ($k>0$), the action must recover
the well-tested Einstein-Hilbert (EH) action. But at the cosmic scale
($k\rightarrow0$), we know that the EH action deviates from observations
and the cosmological constant becomes important. In this picture,
as $k\rightarrow0$, the action leaving $2\lambda+\lambda\nu$ should
play the role of the standard EH action with a limit constant background
scalar curvature $R_{0}$ plus the cosmological constant, so
\begin{equation}
2\lambda+\lambda\nu=\frac{R_{0}-2\Lambda}{16\pi G}.
\end{equation}
While at $k\rightarrow\infty$, $\lambda\tilde{N}\rightarrow\nu$,
the action leaving only the fiducial Lagrangian $\frac{D}{2}\lambda=2\lambda$
which should be interpreted as a constant EH action without the cosmological
constant
\begin{equation}
2\lambda=\frac{R_{0}}{16\pi G}.
\end{equation}
Thus we have the cosmological term 
\begin{equation}
\lambda\nu=\frac{-2\Lambda}{16\pi G}=-\rho_{\Lambda}.
\end{equation}
The action can be rewritten as an effective EH action plus a cosmological
term
\begin{equation}
S_{eff}=\int_{M^{4}}dV\sqrt{|g_{k}|}\left(\frac{R_{k}}{16\pi G}+\lambda\nu\right)\quad(\textrm{small\,k}),
\end{equation}
where
\begin{equation}
\frac{R_{k}}{16\pi G}=2\lambda-\frac{R_{0}}{64\pi^{2}}k^{2},\label{eq:eff-R}
\end{equation}
which is nothing but the flow equation of the scalar curvature \citep{topping2006lectures}
\begin{equation}
R_{k}=\frac{R_{0}}{1+\frac{1}{4\pi}Gk^{2}},\quad\textrm{or}\quad R_{\tau}=\frac{R_{0}}{1+\frac{2}{D}R_{0}\tau}.
\end{equation}
Since at the cosmic scale $k\rightarrow0$, the effective scalar curvature
is bounded by $R_{0}$ which can be measured by ``Hubble's constant''
$H_{0}$ at the cosmic scale,
\begin{equation}
R_{0}=D(D-1)H_{0}^{2}=12H_{0}^{2},
\end{equation}
so $\lambda$ is nothing but the critical density of the 4-spacetime
Universe
\begin{equation}
\lambda=\frac{3H_{0}^{2}}{8\pi G}=\rho_{c},\label{eq:critical density}
\end{equation}
so the cosmological constant is always of order of the critical density
with a ``dark energy'' fraction
\begin{equation}
\Omega_{\Lambda}=\frac{\rho_{\Lambda}}{\rho_{c}}=-\nu\approx0.8,
\end{equation}
which is close to the observational value. The detail discussions
about the cosmological constant problem and the observational effect
in the cosmology, especially the modification of the Distance-Redshift
relation leading to the acceleration parameter $q_{0}\approx-0.68$
can be found in \citep{Luo2015Dark,Luo:2015pca,Luo:2019iby,Luo:2021zpi}. 

If matters are incorporated into the gravity theory, consider the
entangled system in $\mathcal{H}_{\psi}\otimes\mathcal{H}_{X}$ between
the to-be-studied quantum system (matters) and the quantum reference
frame fields system (gravity). $2\lambda$ term in eq.(\ref{eq:coupled gravity (1)})
is normalized by the Ricci flow, by using eq.(\ref{eq:eff-EH+cc})
and eq.(\ref{eq:eff-R}), a matter-coupled-gravity is emerged from
the Ricci flow
\begin{align}
S[\psi,X]\overset{(2)}{\approx} & \int dV\sqrt{|g_{k}|}\left[\frac{1}{2}g^{\mu\nu}\frac{\delta\psi}{\delta X^{\mu}}\frac{\delta\psi}{\delta X^{\nu}}-V(\psi)+2\lambda-\frac{R_{0}}{64\pi^{2}}k^{2}+\lambda\nu\right]\nonumber \\
= & \int dV\sqrt{|g_{k}|}\left[\frac{1}{2}g^{\mu\nu}\frac{\delta\psi}{\delta X^{\mu}}\frac{\delta\psi}{\delta X^{\nu}}-V(\psi)+\frac{R_{k}}{16\pi G}+\lambda\nu\right]
\end{align}

\section{Thermal Equilibrium State}

A Gradient Shrinking Ricci Soliton (GSRS) configuration as a Ricci
flow limit extremizes the Shannon entropy $N$ and the W-functional.
Similarly, a thermal equilibrium state also extremizes the H-functional
of Boltzmann and the thermodynamic entropy. Thus the process of a
generic Ricci flow flows into a GSRS limit is in analogy with the
non-equilibrium state evolves into a thermal equilibrium state, they
are not merely similar but even equivalent, when the thermal system
is nothing but the frame fields system. In this section, following
the previous discussions on the non-equilibrium state of the frame
fields in 4-dimension, in a proper choice of time, we will discuss
the thermal equilibrium state of the frame particle system as a GSRS
configuration in lower 3-dimension, in which the temperature and several
thermodynamic functions of the system can be explicitly calculated
and the manifolds density can be interpreted as the thermal ensemble
density of the frame fields particles, giving a statistical interpretation
to Perelman's thermodynamic analogies of the Ricci flow. 

\subsection{A Temporal Static Shrinking Ricci Soliton as a Thermal Equilibrium
State}

When the shrinking Ricci soliton $M^{4}$ is static in the temporal
direction, i.e. being a product manifolds $M^{4}=M^{3}\times\mathbb{R}$
and $\delta\mathbf{X}/\delta X_{0}=0$, where $X_{0}\in\mathbb{R}$
is the physical time, $\mathbf{X}=(X_{1},X_{2},X_{3})\in M^{3}$ is
a 3-space gradient shrinking Ricci soliton of lower dimensions, we
can prove here that the temporal static spatial part $M^{3}$ is in
thermal equilibrium with the flow parameter $\tau$ proportional to
its temperature, and the manifolds density $u$ of $M^{3}$ can be
interpreted as the thermal equilibrium ensemble density. 

According to Masubara's formalism of thermal fields theory, the thermal
equilibrium of the spatial frame fields can be defined by the periodicity
$\mathbf{X}(\mathbf{x},0)=\mathbf{X}(\mathbf{x},\beta)$ in their
Euclidean time of the lab (remind that we start from the Euclidean
base space for the frame fields theory), where $\beta=1/T$ is the
inverse of the temperature. Now the frame fields is a mapping $\mathbb{R}^{3}\times S^{1}\rightarrow M^{3}\times\mathbb{R}$.
Then in such configuration, the $\tau$ parameter of the 3-space shrinking
soliton $M^{3}$ becomes
\begin{equation}
\tau=\frac{1}{2\lambda}\int\frac{d^{3}\mathbf{p}d\omega_{n}}{(2\pi)^{4}}\frac{1}{\mathbf{p}^{2}+\omega_{n}^{2}}=\frac{1}{2\lambda}T\sum_{n}\int\frac{d^{3}\mathbf{p}}{(2\pi)^{3}}\frac{1}{\mathbf{p}^{2}+(2\pi nT)^{2}},
\end{equation}
where $\omega_{n}=2\pi nT$, $\int\frac{d\omega_{n}}{2\pi}=T\sum_{n}$
have been used. The calculation is a periodic-Euclidean-time version
of the general eq.(\ref{eq:sigma=00003Dtau*g}). Since the density
matrix eq.(\ref{eq:u*}) of the frame fields $X_{\mu}$ is Gaussian
or a coherent state, which the oscillators are almost condensed in
the central peak, thus $\omega_{0}=0$ dominants the Masubara sum,
\begin{equation}
\tau=\frac{1}{2\lambda}T\int\frac{d^{3}\mathbf{p}}{(2\pi)^{3}}\frac{1}{\mathbf{p}^{2}}.
\end{equation}

Different from the naive notion of ``temporal static'' at the classical
level, which means w.r.t. the physical clock $X_{0}$ of the quantum
reference frame, i.e. $\langle\frac{\delta\mathbf{X}}{\delta X_{0}}\rangle=0$.
However, the notion ``temporal static'' is a little subtle at the
quantum level. Because there is no ``absolute static'' at the quantum
or microscopic level, since at such microscopic scale the modes are
always in motion or vibrating w.r.t. the infinitely precise lab time
$x_{0}$, i.e. $\frac{\partial\mathbf{X}(x)}{\partial x_{0}}\neq0$.
Actually $\partial\mathbf{X}/\partial x_{0}$ is in general non-zero
even though its oscillation degrees of freedom are almost frozen (Masubara
frequency $\omega_{n}$ is zero for the Gaussian wavefunction), while
the center of the Gaussian wave pocket of $\mathbf{X}$ is in translational
motion so $\mathbf{p}\neq0$, so its expectation value is in general
finite, for instance, $\langle\frac{\partial\mathbf{X}(x)}{\partial x_{0}}\rangle\sim\frac{3}{2}T<\infty$
claimed by the equipartition energy of the translational motion in
3-space. In general, whether or not the modes of the spatial frame
fields is temporal static depends on the scale to evaluate the average
of the physical clock $\langle X_{0}\rangle$. The notion of ``thermal
static'' in the sense of statistical physics is approximate at a
macroscopic scale rather than a microscopic scale, at which scale
the molecules are always in motion (so does the physical clock $X_{0}$).
The macroscopic scale of the thermal static system is at such a long
physical time scale $\delta\langle X_{0}\rangle\gg\delta x_{0}$ that
the averaged physical clock is almost frozen $\frac{\partial x_{0}}{\partial\langle X_{0}\rangle}\rightarrow0$
w.r.t. the infinitely precise lab time $x_{0}$, so that the thermal
static condition $\langle\frac{\delta\mathbf{X}}{\delta X_{0}}\rangle=\langle\frac{\partial\mathbf{X}}{\partial x_{0}}\rangle\cdot\frac{\partial x_{0}}{\partial\langle X_{0}\rangle}\rightarrow0$
can be achieved. 

More precisely, when we mention that the 3-space is macroscopic ``temporal
static'', a IR cutoff, for example, $H_{0}$ as a macroscopic Hubble
scale should be taken into account. The fluctuation modes on the 3-space
outside the Hubble scale $0<|\mathbf{p}|<H_{0}$ are frozen and temporal
static, while those modes $|\mathbf{p}|>H_{0}$ inside the Hubble
horizon are dynamic. So with this cutoff scale we have 
\begin{equation}
\tau=\frac{1}{2\lambda}T\int_{0}^{|\mathbf{p}|=H_{0}}\frac{d^{3}\mathbf{p}}{(2\pi)}\frac{1}{\mathbf{p}^{2}}=\frac{C_{3}}{2\lambda}TH_{0}=\frac{1}{\lambda_{3}}T=\frac{1}{\lambda_{3}\beta},\label{eq:tau-T}
\end{equation}
where the 3-space energy density is $\lambda_{3}=\frac{\lambda}{\frac{1}{2}C_{3}H_{0}}=\frac{12\pi^{2}\lambda}{H_{0}}$.
Note that if we consider the temporal integral is also cutoff at about
a long physical time scale, e.g. the age of the universe $O(1/H_{0})$,
let the temporal direction is normalized as $\frac{1}{12\pi^{2}}\int_{0}^{12\pi^{2}/H_{0}}dx_{0}H_{0}=1$,
then the condition $\int d^{4}x\lambda\equiv1$ gives its 3-space
version 
\begin{equation}
\int d^{3}x\lambda_{3}=1,\label{eq:3-density normalize}
\end{equation}
which is the definition of $\lambda_{3}$ on 3-space slice generalizing
the critical density $\lambda$ in a 4-spacetime covariant theory. 

It is worth stressing that since the spatial slice depends on the
definition of time, so the value of $\lambda_{3}$ is not universal
(not necessarily equal to above $\frac{12\pi^{2}\lambda}{H_{0}}$
in other frame or cutoff, unlike the universal 4-spacetime critical
density $\lambda$) but frame dependent. If a specific gauge of time
or frame is chosen, $\lambda_{3}$ could be considered fix and be
used as a proportional to correlate the $\tau$ parameter with the
temperature of the temporal static frame fields configuration in such
a specific gauge of time. The 3-space energy density $\lambda_{3}$
is very useful when we consider a temporal static GSRS spacetime or
corresponding thermal equilibrium frame fields ensemble in later discussions. 

In summary, an important observation is that when $M^{3}$ is a shrinking
Ricci soliton in a temporal static product shrinking soliton $M^{3}\times\mathbb{R}$,
the global $\tau$ parameter of $M^{3}$ can be interpreted as a thermal
equilibrium temperature defined by the Euclidean time periodic of
the frame fields, up to a proportional being a 3-space energy density
$\lambda_{3}$ (satisfying eq.(\ref{eq:3-density normalize})) balancing
the dimensions between $\tau$ and $T$. Since temperature $T$ is
frame dependent, so is the proportional $\lambda_{3}$. The observation
gives us a reason why in Perelman's paper $\tau$ could be analogous
to the temperature $T$. The same results can also be obtained if
one use the Lorentzian signature for the lab or base spacetime of
the frame fields theory (\ref{eq:NLSM}). In this case the thermal
equilibrium of the spatial frame fields instead are subject to periodicity
in the imaginary Minkowskian time $\mathbf{X}(\mathbf{x},0)=\mathbf{X}(\mathbf{x},i\beta)$,
but even though the base spacetime is Wick rotated, the path integral
does not pick any imaginary $i$ factor in front of the action in
(\ref{eq:partition of NLSM}) as the starting point, so the main results
of the discussions retain independent to the signature of the base
spacetime. 

\subsection{Thermodynamic Functions}

For the thermodynamic interpretation of the quantum reference frame
and gravity theory, in this subsection, we derive other thermodynamic
functions of the system beside the temperature in the previous subsection,
which are similar with the ideal gas. So the frame fields system in
the Gaussian approximation can be seen as a system of frame fields
gas, which manifests a underlying statistic picture of Perelman's
thermodynamics analogies of his functionals. As convention, we all
take the temperature $T=\lambda_{3}\tau$, eq.(\ref{eq:tau-T}), $D=3$
and $\lambda$ replaced by $\lambda_{3}$, it is equivalent to choose
a specific gauge of time for the thermal equilibrium frame fields
configuration.

When the spatial shrinking soliton $M^{3}$ is in temporal static
$dX_{0}=0$ and in thermal equilibrium, the partition function of
the thermal ensemble of the frame fields $\mathbf{X}$ can be given
by the trace/integration of the density matrix,
\begin{equation}
Z_{*}(\tau)=\lambda_{3}\int d^{3}\mathbf{X}u(\mathbf{X})=\lambda_{3}\int d^{3}\mathbf{X}e^{-\frac{\mathbf{X}^{2}}{4\tau}}=\lambda_{3}(4\pi\tau)^{3/2},\label{eq:Z*}
\end{equation}
the normalized $u$ density can be given by the 3-dimensional version
of eq.(\ref{eq:u*})
\begin{equation}
u_{*}(\mathbf{X})=\frac{1}{Z_{*}}u(\mathbf{X})=\frac{1}{\lambda_{3}(4\pi\tau)^{3/2}}e^{-\frac{\mathbf{X}^{2}}{4\tau}}.
\end{equation}
The partition function can also be consistently given by (\ref{eq:frame-partition})
with $D=3$ in thermal equilibrium and hence the partition function
of the frame fields in the shrinking soliton configuration
\begin{equation}
Z_{*}(\tau)=e^{\lambda_{3}N_{*}(M^{3})-\frac{3}{2}}=\exp\left[-\lambda_{3}\int_{M^{3}}d^{3}Xu_{*}\log u_{*}-\frac{3}{2}\right]=\lambda_{3}(4\pi\tau)^{3/2}=V_{3}\left(\frac{4\pi\lambda_{3}^{1/3}}{\beta}\right)^{3/2}=Z_{*}(\beta),\label{eq:thermo partition function}
\end{equation}
where $V_{3}=\int d^{3}x$ is the 3-volume with the constraint $\lambda_{3}V_{3}=1$.
The partition function is identified with the partition function of
the canonical ensemble of ideal gas (i.e. non-interacting frame fields
gas in the lab) of temperature $1/\beta$ and gas particle mass $\lambda_{3}^{1/3}$.
The interactions are effectively absorbed into the broadening of the
density matrix and normalized mass of the frame fields gas particles. 

The physical picture of frame fields gas in thermal equilibrium lays
a statistical and physical foundation to Perelman's analogies between
his functionals and thermodynamics equations as follows.

The internal energy of the frame fields gas can be given similar to
the standard internal energy of ideal gas $\frac{3}{2}T$ given by
the equipartition energy of translational motion in 3-space. Consider
$\beta$ as the Euclidean time of the flat lab, the internal energy
seen from an observer in the lab is 
\begin{equation}
E_{*}=-\frac{\partial\log Z_{*}}{\partial\beta}=\lambda_{3}^{2}\tau^{2}\frac{\partial N_{*}}{\partial\tau}=\lambda_{3}^{2}\tau^{2}\mathcal{F}_{*}=\frac{3}{2}\lambda_{3}\tau=\frac{3}{2}T,\label{eq:internal energy}
\end{equation}
in which (\ref{eq:F*}) with $D=3$ and $\lambda\rightarrow\lambda_{3}$
have been used. 

The fluctuation of the internal energy is given by
\begin{equation}
\left\langle E_{*}^{2}\right\rangle -\left\langle E_{*}\right\rangle ^{2}=\frac{\partial^{2}\log Z_{*}}{\partial\beta^{2}}=\frac{3}{2}\lambda_{3}^{2}\tau^{2}=\frac{3}{2}T^{2}.
\end{equation}
The Fourier transformation of the density $u_{*}(\mathbf{X})$ is
given by
\begin{equation}
u_{*}(\mathbf{K})=\int d^{3}Xu_{*}(\mathbf{X})e^{-i\mathbf{K}\cdot\mathbf{X}}=e^{-\tau\mathbf{K}^{2}},
\end{equation}
since $u$ satisfies the conjugate heat equation (\ref{eq:u-equation}),
so $\mathbf{K}^{2}$ is the eigenvalue of the Laplacian $-4\Delta_{X}+R$
of the 3-space, taking the value of the F-functional,
\begin{equation}
\mathbf{K}^{2}=\lambda_{3}\int d^{3}X\left(R|\Psi|^{2}+4|\nabla\Psi|^{2}\right)=\lambda_{3}\mathcal{F},
\end{equation}
so
\begin{equation}
u_{*}(\mathbf{K}^{2})=e^{-\lambda_{3}\tau\mathcal{F}}.
\end{equation}
For a state taking energy $\lambda_{3}^{2}\tau^{2}\mathcal{F}=E$,
the probability density of the state can be rewritten as
\begin{equation}
u_{*}(E)=e^{-\frac{E}{\lambda_{3}\tau}}=e^{-\frac{E}{T}},\label{eq:ensemble density}
\end{equation}
which is the standard Boltzmann's probability distribution of the
state. So we can see that the (Fourier transformed) manifolds density
can be interpreted as the thermal equilibrium canonical ensemble density
of the frame fields.

The free energy is given by
\begin{equation}
F_{*}=-\frac{1}{\beta}\log Z_{*}=-\lambda_{3}\tau\log Z_{*}=-\frac{3}{2}\lambda_{3}\tau\log(4\pi\tau),\label{eq:free-energy}
\end{equation}
similar with the standard free energy of ideal gas $-\frac{3}{2}T\log T$
up to a constant.

The minus H-functional of Boltzmann at an equilibrium limit and the
thermal entropy of the frame fields gas can be given by the Shannon
entropy
\begin{equation}
\lambda_{3}N_{*}=S_{*}=-\lambda_{3}\int d^{3}Xu_{*}\log u_{*}=\frac{3}{2}\left[1+\log(4\pi\tau)\right],\label{eq:S*}
\end{equation}
similar with the thermal entropy of fixed-volume ideal gas $\frac{3}{2}\log T+\frac{3}{2}$
up to a constant. The thermal entropy can also be consistently given
by the standard formula
\begin{equation}
S_{*}=\log Z_{*}-\beta\frac{\partial\log Z_{*}}{\partial\beta}=\frac{3}{2}\left[1+\log(4\pi\tau)\right].\label{eq:bulk entropy}
\end{equation}
which is in analogy with the fact that the W functional is the Legendre
transformation of the relative Shannon entropy w.r.t. $\tau^{-1}$.
For this reason, the W functional is also an entropy function related
to the (minus) thermodynamics entropy.

In summary, we have seen that, under general frame fields (coordinates)
transformation the Shannon entropy anomaly $N$ appearing in the partition
function (\ref{eq:Z->Zhat}) (or relative Shannon entropy $\tilde{N}$
w.r.t. $N_{*}$) has profound thermodynamics interpretations. The
Ricci flow of frame fields lead to non-equilibrium and equilibrium
thermodynamics of the quantum spacetime, we summarize the comparisons
between them in the Table I and II.

\begin{table}[H]
\centering{}%
\begin{tabular}{|c|c|}
\hline 
Frame fields at non-Ricci-flow-limit & Non-equilibrium thermodynamics\tabularnewline
\hline 
\hline 
Relative Shannon entropy: $\tilde{N}=-\int d^{3}\mathbf{X}\tilde{u}(\mathbf{X},t)\log\tilde{u}(\mathbf{X},t)$ & Boltzmanian H function: $H(t)=\int d^{3}\mathbf{v}\rho(\mathbf{v},t)\log\rho(\mathbf{v},t)$\tabularnewline
\hline 
Ricci flow parameter: $t$ & Newtonian time: $t$\tabularnewline
\hline 
Monotonicity: $\frac{d\tilde{N}}{dt}=-\tilde{\mathcal{F}}\ge0$ & H theorem: $\frac{dH}{dt}\le0$\tabularnewline
\hline 
conjugate heat equation: $\frac{\partial u}{\partial t}=\left(-\varDelta+R\right)u$ & Boltzmann equation of ideal gas: $\frac{\partial\rho}{\partial t}=-\mathbf{v}\cdot\boldsymbol{\nabla}\rho$\tabularnewline
\hline 
\end{tabular}\caption{Frame fields in general Ricci flow at non-flow-limit and the Non-equilibrium
thermodynamics.}
\end{table}

\begin{table}[H]

\begin{centering}
\begin{tabular}{|c|c|}
\hline 
Frame fields at the Ricci flow limit (GSRS) & Equilibrium thermodynamics of ideal gas\tabularnewline
\hline 
\hline 
partition function: $Z_{*}(\tau)=\lambda_{3}(4\pi\tau)^{3/2}$ & partition function: $Z(T)=V_{3}(2\pi mT)^{3/2}$\tabularnewline
\hline 
GSRS flow parameter: $\lambda_{3}\tau$ & temperature: $T=\beta^{-1}$\tabularnewline
\hline 
$\lambda_{3}^{2}\tau^{2}\mathcal{F}_{*}=\frac{3}{2}\lambda_{3}\tau$ & internal energy: $E_{*}=-\frac{\partial\log Z}{\partial\beta}=\frac{3}{2}T$\tabularnewline
\hline 
manifold density: $u_{*}(\mathbf{K})=e^{-\tau\mathbf{K}^{2}}=e^{-\lambda_{3}\tau\mathcal{F}}$ & canonical ensemble density: $\rho=e^{-\frac{E}{T}}$\tabularnewline
\hline 
$-\lambda_{3}\tau\log Z_{*}=-\frac{3}{2}\lambda_{3}\tau\log(4\pi\tau)$ & free energy: $F_{*}=-T\log Z(T)=-\frac{3}{2}T\log T$\tabularnewline
\hline 
Shannon entropy: $\lambda_{3}N_{*}=\frac{3}{2}\left[1+\log(4\pi\tau)\right]$ & thermodynamic entropy: $S_{*}=\frac{3}{2}\left(1+\log T\right)$\tabularnewline
\hline 
W functional: $\mathcal{W}=\tau\frac{d\tilde{N}}{d\tau}+\tilde{N}$ & first law of thermodynamics: $E_{*}-TS_{*}=F_{*}$\tabularnewline
\hline 
Monotonicity: $\frac{d\tilde{N}}{dt}\ge0$ & second law of thermodynamics: $\delta S\ge0$\tabularnewline
\hline 
\end{tabular}\caption{Frame fields in Gradient Shrinking Ricci Soliton (GSRS) configuration
and the equilibrium thermodynamics of ideal gas.}
\par\end{centering}
\end{table}

\section{Application to the Schwarzschild Black Hole}

In this section, we try to apply the general statistic and thermodynamics
interpretation of the quantum frame fields to a physical gravitational
system, as one of the touchstone of quantum gravity, i.e. to understand
the statistical origin of the thermodynamics of the Schwarzschild
black hole.

\subsection{The Temperature of a Schwarzschild Black Hole}

The region in the vicinity of the origin of a Schwarzschild black
hole is an example of classical static shrinking Ricci soliton. A
rest observer distant from it sees an approximate metric $M^{3}\times\mathbb{R}$,
where the region in the vicinity of the origin of the spatial part
$M^{3}$ is a shrinking Ricci soliton. The reason is as follows, because
the black hole satisfies the Einstein's equation
\begin{equation}
R_{\mu\nu}-\frac{1}{2}g_{\mu\nu}R=8\pi G\mathcal{T}_{\mu\nu},
\end{equation}
where the stress tensor is a point distributed matter in rest with
a mass $m$ at the origin $x=0$ (seen from the distant rest observer)
\begin{equation}
\mathcal{T}_{00}=m\delta^{(3)}(\mathbf{x}),\quad\mathcal{T}_{ij}=0\quad(i,j=1,2,3),
\end{equation}
where Latin index $i,j$ is for spatial index in the following. So
we have 
\begin{equation}
R(\mathbf{x})=-8\pi G\mathcal{T}_{\mu}^{\mu}=8\pi Gm\delta^{(3)}(\mathbf{x}).\label{eq:scalar curvature of bh}
\end{equation}
From the Einstein's equation we have the Ricci curvature of $M^{3}$
is proportional to the metric of $M^{3}$ 
\begin{equation}
R_{ij}(\mathbf{x})=8\pi G\mathcal{T}_{ij}+\frac{1}{2}g_{ij}R=\frac{1}{2}8\pi Gm\delta^{(3)}(\mathbf{x})g_{ij}\quad(i,j=1,2,3).
\end{equation}
The equation is nothing but a normalized shrinking Ricci soliton equation
(\ref{eq:shrinker}) for $M^{3}$
\begin{equation}
R_{ij}(\mathbf{x})=\frac{1}{2\tau}g_{ij}(\mathbf{x})\quad(\mathbf{x}\approx0)\label{eq:3-shrinking-soliton}
\end{equation}
with 
\begin{equation}
\delta^{(3)}(\mathbf{x})\tau=\frac{1}{8\pi Gm},
\end{equation}
where $\delta^{(3)}(x)$ plays the role of the 3-space energy density
$\lambda_{3}$ in the vicinity of the origin, satisfying $\int d^{3}x\delta^{(3)}(x)=1$
as eq.(\ref{eq:3-density normalize}), so by using the relation between
$\tau$ and temperature $T$ (\ref{eq:tau-T}), we can directly read
from the equation that a temperature seen by the lab's infinite distant
rest observer is
\begin{equation}
T=\delta^{(3)}(\mathbf{x})\tau=\frac{1}{8\pi Gm},\label{eq:BH temperature}
\end{equation}
which is the standard Hawking's temperature of the Schwarzschild black
hole seen by a distant rest observer.

Is the vacuum region outside the origin of the black hole also a shrinking
Ricci soliton? One may naively think that the answer is no, since
at the classical level, it seems $R_{ij}=0$ (not a shrinking soliton
eq.(\ref{eq:3-shrinking-soliton})), since outside the origin is just
vacuum. But as is discussed in the next subsection, we argue that
it is not true at the quantum level, if the vacuum and the vicinity
region of the origin are in thermal equilibrium, they must be a shrinking
Ricci solitons as a whole, i.e. $\langle R_{ij}\rangle=\frac{1}{2\tau}g_{ij}\neq0$,
eq.(\ref{eq:<Rij>=00003D(1/2tau)*gij}) in the ``vacuum''. The above
result can be extended to the ``vacuum'' region outside the origin,
the price to pay is that the ``vacuum'' is full of internal energy
corresponding to the Hawking temperature. If the whole spacetime have
not been in thermal equilibrium yet, the configuration has to irreversibly
go on flowing to a common thermal equilibrium fixed point (a global
shrinking Ricci soliton), leading to a global maximized entropy, as
the H-theorem asserts.

\subsection{The Energy of a Schwarzschild Black Hole}

In classical general relativity, the mass $m$ is often mentioned
as the ADM energy of the black hole
\begin{equation}
m=\int d^{3}\mathbf{x}\mathcal{T}_{00}=\int d^{3}\mathbf{x}m\delta^{(3)}(\mathbf{x}),
\end{equation}
seen by the distant rest observer (w.r.t. the lab time $x_{0}$).
Here at the quantum level, the coordinates or frame fields and spacetime
are quantum fluctuating, which gives rise to the internal energy related
to the periodicity of the (Euclidean) lab time $x_{0}$ (i.e. $\beta=\frac{1}{T}$).
So, mathematically speaking, the anomaly of the trace of the stress
tensor will modify the total ADM mass at the quantum level, see (\ref{eq:trace anomaly}).
Since the anomaly of the action of the frame fields $\lambda_{3}N_{*}$
representing the spacetime part is always real, the internal energy
of the frame fields is given by the (\ref{eq:internal energy})
\begin{equation}
E_{*}=-\frac{\partial\log Z_{*}}{\partial\beta}=\frac{3}{2}T=\frac{3}{16\pi Gm},\label{eq:internal energy of BH}
\end{equation}
in which we have considered the 3-space volume $V_{3}$ outside the
origin is in thermal equilibrium with the Hawking's temperature at
the origin eq.(\ref{eq:BH temperature}), sharing the same equilibrium
temperature $T$ in the 3-volume $V_{3}$. 

We can see that the internal energy $E_{*}$ is an extra contribution
to the total energy of the (black hole + ``vacuum'') system seen
by the distant rest observer. Essentially this term can be seen as
a quantum correction or a part of the trace anomaly contribution to
the stress tensor, thus the total energy of the black hole including
the classical ADM energy and the quantum fluctuating internal energy
of the metric is
\begin{equation}
m_{BH}=\int d^{3}\mathbf{x}\left\langle \mathcal{T}_{00}\right\rangle =m+E_{*}=m+\frac{3}{16\pi G}\frac{1}{m},\label{eq:BH total mass}
\end{equation}
where the classical stress tensor $\mathcal{T}_{00}$ is formally
replaced by its quantum expectation value
\begin{equation}
\langle\mathcal{T}_{00}\rangle=m\delta^{(3)}(\mathbf{x})+\frac{3}{2}\frac{T}{V_{3}}.\label{eq:<Tij>}
\end{equation}
A quantum Equivalence Principle should assert that the total energy
rather than only the classical ADM mass contributes to the gravitation.

For a macroscopic classical black hole, $m\gg\sqrt{\frac{1}{G}}$,
the first term ADM energy dominants the eq.(\ref{eq:BH total mass}),
\begin{equation}
m_{BH}\approx m.
\end{equation}
The second internal energy term is gradually non-negligible for a
microscopic quantum black hole. An important effect of the existence
of the second term in (\ref{eq:BH total mass}) is, for a microscopic
quantum black hole, it makes the total energy bound from below, the
minimal energy is of order of the Planck mass
\begin{equation}
m_{BH}\ge\sqrt{\frac{3}{4\pi G}}\sim O(m_{p}),
\end{equation}
which seems to prevent the black hole evaporating into nothing.

Further more, the internal energy $\frac{3}{2}T$ term contributing
to the total energy $m_{BH}$ and gravitation also demands that, not
only the vicinity of the origin of the black hole is a shrinking soliton
(as previous subsection claims), at the quantum level the whole 3-space
is also the same shrinking soliton (i.e. satisfying eq.(\ref{eq:3-shrinking-soliton})
with the identical $\tau$ globally and hence the same temperature
$T$ everywhere for the whole 3-space), just replacing the $\delta^{(3)}$-density
in eq.(\ref{eq:BH temperature}) by the $\lambda_{3}$-density, which
extends the $\delta^{(3)}$-density at the origin to the outside region
(the ``vacuum''), we have
\begin{equation}
T=\lambda_{3}\tau=\frac{1}{8\pi Gm},\quad\mathrm{with}\quad\int d^{3}\mathbf{x}\lambda_{3}=\int d^{3}\mathbf{x}\frac{\left\langle \mathcal{T}_{00}\right\rangle }{m_{BH}}=1
\end{equation}
for the whole thermal equilibrium 3-space, although at the classical
level the vacuum $R_{ij}(x\neq0)=0$ is seem not a shrinking soliton
outside the origin. The physical reason is transparent that the internal
energy's contribution $\frac{3T}{2V_{3}}$ in $\langle\mathcal{T}_{00}\rangle$
also plays the role of an additional source of gravity outside the
origin. For the whole 3-space with $\langle\mathcal{T}_{00}\rangle\neq0$
and $\langle\mathcal{T}_{ij}\rangle=0$, the Einstein's equation for
the whole 3-space is nothing but the Shrinking Ricci Soliton equation
(\ref{eq:3-shrinking-soliton}):
\begin{equation}
\langle R_{ij}\rangle=\frac{1}{2}\left\langle R\right\rangle g_{ij}=\frac{1}{2}8\pi G\left\langle \mathcal{T}_{00}\right\rangle g_{ij}\approx\frac{1}{2}8\pi Gm\frac{\left\langle \mathcal{T}_{00}\right\rangle }{m_{BH}}g_{ij}=\frac{1}{2T}\lambda_{3}g_{ij}=\frac{1}{2\tau}g_{ij}\neq0\label{eq:<Rij>=00003D(1/2tau)*gij}
\end{equation}
in which $\langle R\rangle=-8\pi G\langle\mathcal{T}_{\mu}^{\mu}\rangle=8\pi G\langle\mathcal{T}_{00}\rangle\neq0$
is used in the ``vacuum'' outside the origin. The equation is in
fact the spatial components of the Gradient Shrinking Ricci Soliton
equation (\ref{eq:shrinker}) where $\langle R_{ij}\rangle=R_{ij}+\nabla_{i}\nabla_{j}f$,
the Gaussian/thermal broadening of the density matrix $u$ contributes
to the classical curvature. The vicinity region of the origin plus
the ``vacuum'' outside the origin of the black hole as a whole,
is nothing but globally a shrinking Ricci soliton. The \textquotedblleft vacuum\textquotedblright{}
is not completely nothing at the quantum level but full of thermal
particles $\langle\mathcal{T}_{00}(x\neq0)\rangle\neq0$. The Hawking
temperature is essentially an Unruh effect, in certain sense, the
Gradient Shrinking Ricci soliton equation, eq.(\ref{eq:<Rij>=00003D(1/2tau)*gij}),
might play a more fundamental role than the Unruh's formula, which
determines how local acceleration or gravitation gives rise to temperature.

The internal energy of the spacetime frame fields is an additional
and necessary source of gravity, although macroscopically it is too
small to contribute, at the quantum level its contribution is crucial
for the 3-space in thermal equilibrium just right being a global shrinking
Ricci soliton. The thermal internal energy coming from the quantum
fluctuation of the 3-space gravitates normally as the quantum Equivalence
Principle will assert. Otherwise, we have to face a paradox as follows.
If we consider a frame $x$ having $\mathcal{T}_{\mu\nu}(x)=0$ everywhere,
so according to the classical gravity $R_{\mu\nu}(x)=0$ everywhere,
if we transform it to anther accelerating frame $x^{\prime}$, one
expects $\mathcal{T}_{\mu\nu}(x)\rightarrow\mathcal{T}_{\mu\nu}^{\prime}(x^{\prime})=0$,
and hence $R_{\mu\nu}^{\prime}(x^{\prime})=0$ everywhere. However,
according to the Equivalence Principle, in the accelerating frame
$x^{\prime}$ one should feel equivalent gravity $R_{\mu\nu}^{\prime}(x^{\prime})\neq0$.
It is clearly something is missing, a new dimension of the Equivalence
Principle must be considered. In order to solve the paradox and retain
the Equivalence Principle, a quantum effect (actually the effect from
the diffemorphism anomaly such as the trace anomaly or the Unruh effect)
must be introduced so that the accelerating frame must be particles
creating from the ``vacuum'' and be thermal, which plays the role
of an equivalent gravitational source making $R_{\mu\nu}^{\prime}(x^{\prime})\neq0$.
The Hawking temperature in the internal energy term of eq.(\ref{eq:BH total mass})
is in essential the Unruh temperature playing such role. In this sense,
the validity of the Equivalence Principle should be extended to the
reference frame described by quantum state.

\subsection{The Entropy of a Schwarzschild Black Hole}

In the general framework, the entropy of the black hole comes from
the uncertainty or quantum fluctuation moment of the frame fields
given by the manifolds density $u$, more precisely, the thermalized
black hole entropy is measured by the maximized Shannon entropy in
terms of the probability distribution $u$ of the frame fields in
the background of the black hole. So in this subsection, we calculate
the $u$ density distributed around the Schwarzschild black hole and
then evaluate the corresponding entropy as a measure of the black
hole entropy. After a proper definition of a zero-point of the Shannon
entropy, it gives a standard Bekenstein-Hawking entropy. 

For an observer in the distant lab rest frame, the contributions to
the temporal static $u$ density around the black hole is two folds.
Beside the thermal distribution $u_{*}$ in the ``vacuum'' or bulk
outside the black hole horizon, which gives rise to the ideal gas
entropy (\ref{eq:S*}) as the background entropy, there is an additional
$\tilde{u}$ density distribute mostly in a exterior thin shell near
the horizon, and sparsely in the bulk outside the horizon, which we
will focus on. The reason is as follows. Because $\tilde{u}$ density
satisfies the conjugate heat equation (\ref{eq:u-equation}) on the
classical background of the black hole, since the classical scalar
curvature $R=0$ outside the horizon, and the temperature (equivalently
the parameter $\tau$ and the mass) can be seen unchange for the thermalized
black hole i.e. $\frac{\partial\tilde{u}}{\partial\tau}=0$, thus
the conjugate heat equation for $\tilde{u}$ is approximately given
by the 4-Laplacian equation on the Schwarzschild black hole
\begin{equation}
\Delta_{X}\tilde{u}(X)=0,\quad(|\mathbf{X}|\ge r_{H}).
\end{equation}
Now the temporal static density $\tilde{u}(X)$ plays a similar role
like a solution of the Klein-Gordon equation on the static background
of the black hole. The approximation of the conjugate heat equation
is equivalent to interpret the Klein-Gordon modes as a ``first''-quantization
probability density (not second-quantization fields). As is well-known,
there are modes falling into the black hole horizon and hence disappearing
from the outside observer's view. Just like the negative Klein-Gordon
modes falling into the negative energy states below the groundstate.
In a flat background, the amplitudes of the modes falling into and
going out of the horizon are identical. So in the second-quantization,
the negative mode falling into the horizon can be reinterpreted as
a single anti-particle with positive energy modes going out of the
horizon with the identical amplitude. However, in a curved background,
for instance, the spacetime near the black hole horizon, the statement
is no longer true. The two amplitudes differ from each other by a
non-unitary equivalent factor. Thus the negative mode falling into
the black hole horizon are no longer be reinterpreted as a single
anti-particle mode going out, rather than multi-particles thermo-ensemble.
At the situation, the density $\tilde{u}$ describes the ensemble
density of modes going exterior the horizon $|\mathbf{X}|\ge r_{H}$
which can be seen by an outside observer.

By a routine calculation of the solution near the exterior black hole
horizon resembling a Rindler metric as a starting point, we denote
the solution $\tilde{u}_{\mathbf{k}}(\rho)$, in which $\mathbf{k}$
represents the Fourier component/momentum in the direction that are
orthogonal to the direction of radius with $\rho=\log(r-r_{H})$,
$r$ the radius, $r_{H}=2Gm$ the radius of the horizon. The equation
becomes
\begin{equation}
-\frac{\partial^{2}\tilde{u}_{\mathbf{k}}}{\partial\rho^{2}}+\mathbf{k}^{2}e^{2\rho}\tilde{u}_{\mathbf{k}}=\omega^{2}\tilde{u}_{\mathbf{k}},
\end{equation}
where $\omega$ is the eigen-energy of the modes. By using a natural
boundary condition that $\tilde{u}$ vanishes at infinity, we can
see that each transverse Fourier mode $\tilde{u}_{\mathbf{k}}$ can
be considered as a free 1+1 dimensional quantum field confined in
a box, one wall of the box is at the reflecting boundary $\rho_{0}=\log\epsilon_{0}$
where $\epsilon_{0}\approx0$, and the other wall of the box is provided
by the potential 
\begin{equation}
V(\rho)=\mathbf{k}^{2}e^{2\rho},
\end{equation}
which becomes large $V(\rho)\gg1$ at $\rho>-\log\mathbf{k}$. So
we can approximate the potential by the second wall at $\rho_{w}=-\log\mathbf{k}$.
So the length of the box is given by
\begin{equation}
\Delta\rho=\rho_{w}-\rho_{0}=-\log(\epsilon_{0}\mathbf{k}).
\end{equation}
Thus the thickness of the horizon is about $\Delta r\sim e^{\Delta\rho}\sim\epsilon_{0}\mathbf{k}$.

The density $\tilde{u}_{\mathbf{k}}(\rho)$ is located in the box
$\rho\in(\rho_{0},\rho_{w})$. In other words, the solution of $\tilde{u}$
density is located mainly in a thin shell near the horizon $r\in(r_{H},r_{H}+\epsilon_{0}\mathbf{k})$.
Furthermore, the modes $\mathbf{k}$ is assumed normal distributed
(with a tiny width described by the parameter $\tau$). In this picture,
without solving the equation, we can approximately write down the
natural solution as $\tilde{u}_{\mathbf{k}}(r)\overset{\tau\rightarrow0}{\approx}\delta(|\mathbf{k}|)\delta(r-r_{H})$,
while for finite and small $\tau$, we have a nearly Gaussian form
\begin{equation}
\tilde{u}_{\mathbf{k}}(r)\approx\delta(|\mathbf{k}|)\cdot\frac{1}{(4\pi\tau)^{1/2}}e^{-\frac{(r-r_{H})^{2}}{4\tau}}\approx\frac{1}{(4\pi|\mathbf{k}|^{2}\tau)^{1/2}}e^{-\frac{(r-r_{H})^{2}}{4\tau}},\quad(r>r_{H})
\end{equation}
The exterior horizon solution can be considered as the standing wave
solution as the superposition of the modes falling into and coming
out of the black hole horizon. Then we have (up to a constant)
\begin{equation}
\log\tilde{u}_{\mathbf{k}}(r)\overset{r\sim r_{H}}{\approx}-\frac{1}{2}\log\left(|\mathbf{k}|^{2}\tau\right).
\end{equation}
A routine calculation of the relative Shannon entropy or W-functional
gives the entropy of each k-mode in the limit in which the width $\tau$
is very small,
\begin{align}
\lambda_{3}\tilde{N}(\tilde{u}_{\mathbf{k}}) & =-\lambda_{3}\int d^{3}X\tilde{u}_{\mathbf{k}}\log\tilde{u}_{\mathbf{k}}\nonumber \\
 & =\delta(|\mathbf{k}|)\int_{r_{H}}^{\infty}4\pi r^{2}dr\frac{1}{(4\pi\tau)^{1/2}}e^{-\frac{(r-r_{H})^{2}}{4\tau}}\frac{1}{2}\log\left(|\mathbf{k}|^{2}\tau\right)\nonumber \\
 & \overset{\tau\sim0}{=}\delta(|\mathbf{k}|)\frac{1}{4}A\log\left(|\mathbf{k}|^{2}\tau\right),\label{eq:entropy-3-volume-integral}
\end{align}
where $A=4\pi r_{H}^{2}$ is the area of the horizon.

It is naturally to assume the momentum $\mathbf{k}$ in the horizon
shell is homogeneous, 
\begin{equation}
|\mathbf{k}|=|k_{r}|=|k_{\perp}|,\label{eq:momentum in shell}
\end{equation}
where $k_{r}$ is the momentum in the radius direction and $k_{\perp}$
in the transverse directions on the horizon. When we integrate over
all k-modes, we have the total relative Shannon entropy weakly depending
to $\tau$
\begin{align}
\lambda_{3}\tilde{N}(\tilde{u}) & =\lambda_{3}\int d^{3}\mathbf{k}\tilde{N}(\tilde{u}_{\mathbf{k}})\nonumber \\
 & =\frac{1}{4}A\int\frac{d^{2}k_{\perp}}{(2\pi)^{2}}\log\left(|k_{\perp}|^{2}\tau\right)\int dk_{r}\delta(k_{r})\nonumber \\
 & \approx\frac{1}{4}A\int_{0}^{1/\epsilon}\frac{2\pi k_{\perp}dk_{\perp}}{(2\pi)^{2}}\log\left(|k_{\perp}|^{2}\tau\right)\nonumber \\
 & =\frac{1}{4}A\times\frac{1}{2\pi\tau}\left[-\frac{\tau}{2\epsilon^{2}}\left(1-\log\frac{\tau}{\epsilon^{2}}\right)\right]\nonumber \\
 & \approx-\frac{A}{16\pi\epsilon^{2}},\label{eq:entropy-3-momentum-integral}
\end{align}
in which the transverse momentum is effectively cut off at an inverse
of a fundamental UV length scale $\epsilon^{2}$.

The relative Shannon entropy gives an area law of the black hole entropy.
To determine the UV length cutoff $\epsilon^{2}$, we need to consider
the scale at which the relative entropy is defined to be zero (not
only the black hole is locally thermal equilibrium, but also the asymptotic
background spacetime is globally thermal equilibrium), thus we need
to consider the flow of the asymptotic background spacetime. A natural
choice of a thermal equilibrium Ricci flow limit of the background
spacetime (the black hole is embedded) is an asymptotic homogeneous
and isotropic Hubble universe with scalar curvature $R_{0}=D(D-1)H_{0}^{2}=12H_{0}^{2}$
at scale $t_{UV}$ where we could consider and normalize the relative
entropy to be zero (leaving only the background ideal gas entropy),
since there is no information of the local shape distortions in such
GSRS background because of the vanishing of its Weyl curvature, while
the global curvature is non-zero which codes the information of its
global volume shrinking. Under such definition, taking the normalized
Shrinking Ricci soliton equation (\ref{eq:shrinker}) and (\ref{eq:tau}),
we have
\begin{equation}
\tau_{UV}=-t_{UV}=\frac{D}{2R_{0}}=\frac{1}{64\pi^{2}\lambda}k_{UV}^{2},
\end{equation}
by using the critical density (\ref{eq:critical density}), which
gives a natural cutoff corresponding to the scale $t_{UV}$,
\begin{equation}
\epsilon^{2}=k_{UV}^{-2}=\frac{1}{D\pi}G=\frac{1}{4\pi}G.
\end{equation}

This is exactly the Planck scale, which is a natural cutoff scale
induced from the Hubble scale $H_{0}$ and $\lambda$ of the framework.
However, it is worth stressing that the Planck scale is not the absolute
fundamental scale of the theory, it only has meaning w.r.t. the asymptotic
Hubble scale. The only fundamental scale of the theory is the critical
density $\lambda$ which is given by a combination of both the Planck
scale and Hubble scale, but each individual Planck or Hubble scale
does not have absolute meaning. The UV (Planck) cutoff scale could
tend to infinity while the complementary (Hubble) scale correspondingly
tends to zero (asymptotic flat background), keeping $\lambda$ finite
and fixed. 

At this point, if we define a zero-relative-entropy for an asymptotic
Hubble universe of scalar curvature $R_{0}$, then the black hole
in this asymptotic background has a non-zero thermodynamic entropy
\begin{equation}
S=-\lambda_{3}\tilde{N}(\tilde{u})=\frac{A}{4G},\label{eq:S=00003DA/4}
\end{equation}
up to the bulk background entropy $\lambda_{3}N_{*}=S_{*}\ll S$,
eq.(\ref{eq:bulk entropy}). Combining the relative Shannon entropy
$\tilde{N}$ and the bulk thermal background entropy $N_{*}$, and
using the total partition function eq.(\ref{eq:frame-partition}),
$Z(M^{3})=e^{\lambda_{3}N-\frac{3}{2}}=e^{\lambda_{3}(\tilde{N}+N_{*})-\frac{3}{2}}$
we can also reproduce the total energy of the black hole in (\ref{eq:BH total mass})
\begin{equation}
m_{BH}=-\frac{\partial\log Z}{\partial\beta}=m+\frac{3}{2}T,
\end{equation}
in which eq.(\ref{eq:perelman-partition}) and $A=4\pi r_{H}^{2}=16\pi G^{2}m^{2}=\frac{\beta^{2}}{4\pi}$
have been used.

Different from the holographic idea that the information or entropy
are coded in the (infinite thin and 2-dimensional) horizon or boundary
of a gravitational system, in this framework where the coordinates
of the spacetime geometry are smeared by quantum fluctuation, as a
consequence that there is no mathematically precise notion of an infinite
thin boundary in a ``density manifolds'' in general, it is just
a semi-classical concept. Note that manifolds density $u$ is mainly
distributed at the horizon with a finite thickness (although very
small), which contributes most of the anomaly and entropy to the black
hole, so although the entropy (\ref{eq:S=00003DA/4}) is proportional
to the area, the geometric gravitational entropy given by the framework
essentially comes from the 3-volume (note the 3d integral in eq.($\ref{eq:entropy-3-volume-integral}$)
and eq.(\ref{eq:entropy-3-momentum-integral})) but the 2-surface
boundary. Or in other equivalent words, here the area of the horizon
is fluctuating (due to its finite thickness) rather fixed, while the
total energy and hence the temperature is fixed. In this sense, it
is a canonical ensemble but an area ensemble as some ideas might suggest.

\section{Conclusions}

In this paper, we have proposed a statistical fields theory underlying
Perelman's seminal analogies between his geometric functionals and
the thermodynamic functions. The theory is based on a $d=4-\epsilon$
quantum non-linear sigma model, interpreted as a quantum reference
frame. When we quantize the theory at the Gaussian approximation,
the wavefunction $\Psi(X)$ and hence the density matrix $u(X)=\Psi^{*}(X)\Psi(X)$
eq.(\ref{eq:u}) can be written down explicitly. Based on the density
matrix, the Ricci flow of the frame fields (\ref{eq:ricci flow})
and the generalized Ricci-DeTurck flow (\ref{eq:ricci-deturk}) of
the frame fields endowed with the density matrix is discussed. And
further more, we find that the density matrix has profound statistical
and geometric meanings, by using it, the spacetime $(M^{D},g)$ as
the target space of NLSM is generalized to a density spacetime $(M^{D},g,u)$.
The density matrix $u(X,\tau)$, satisfying a conjugate heat equation
(\ref{eq:u-equation}), not only describes a (coarse-grained) probability
density of finding frame fields in a local volume, but also describes
a volume comparison between a local volume and the fiducial one. 

For the non-isometric nature of the Ricci or Ricci-DeTurck flow, the
classical diffeomorphism is broken down at the quantum level. By the
functional integral quantization method, the change of the measure
of the functional integral can be given by using a Shannon entropy
$N$ in terms of the density matrix $u(X,\tau)$. The induced trace
anomaly and its relation to the anomalies in conventional gravity
theories are also discussed. As the Shannon entropy flows monotonically
to its maximal value $N_{*}$ in a limit called Gradient Shrinking
Ricci Soliton (GSRS), a relative density $\tilde{u}$ and relative
Shannon entropy $\tilde{N}=N-N_{*}$ can be defined w.r.t. the flow
limit. The relative Shannon entropy gives a statistical interpretation
underlying Perelman's partition function (\ref{eq:perelman-partition}).
And the monotonicity of $\tilde{N}$ along the Ricci flow gives an
analogous H-theorem (\ref{eq:analog H-theorem}) for the frame fields
system. As a side effect, the meanings on the gravitational side of
the theory is also discussed, in which a cosmological constant $-\lambda\nu(B_{\infty}^{4})\approx0.8\rho_{c}$
as a UV counter term of the anomaly must be introduced.

We find that a temporal static GSRS, $M^{3}$, as a 3-space slice
of the 4-spacetime GSRS, $M^{4}=M^{3}\times\mathbb{R}$, is in a thermal
equilibrium state, in which the temperature is proportional to the
global $\tau$ parameter of $M^{3}$ (\ref{eq:tau-T}) up to a 3-space
energy density $\lambda_{3}$ with normalization $\int d^{3}x\lambda_{3}=1$.
The temperature and $\lambda_{3}$ both depend on the choice of time
$\mathbb{R}$. In the sense that $M^{3}$ is in thermal, its Ricci
soliton equation eq.(\ref{eq:3-shrinking-soliton}) or quantum (indistinguishable
with thermal) fluctuation eq.(\ref{eq:sigma=00003Dtau*g}), can be
considered as a generalization of the Unruh's formula, relating the
temperature to local acceleration or gravitation. Based on the statistical
interpretation of the density matrix $u(X,\tau)$, we find that the
thermodynamic partition function (\ref{eq:Z*}) at the Gaussian approximation
is just a partition function of ideal gas of the frame fields. In
this physical picture of canonical ensemble of frame fields gas, several
thermodynamic functions, including the internal energy (\ref{eq:internal energy}),
the free energy (\ref{eq:free-energy}), the thermodynamic entropy
(\ref{eq:S*}), and the ensemble density (\ref{eq:ensemble density})
etc. can be calculated explicitly agreeing with Perelman's formulae,
which gives an underlying statistical foundation to Perelman's analogous
functionals.

We find that the statistical fields theory of quantum reference frame
can be used to give a possible underlying microscopic origin of the
spacetime thermodynamics. The standard results of the thermodynamics
of the Schwarzschild black hole, including the Hawking temperature,
energy and Bekenstein-Hawking entropy can be successfully reproduced
in the framework. And we find that when the fluctuation internal energy
of the metric is taken into account in the total energy, the energy
of the black hole has a lower bound of order of the Planck energy,
which avoid the quantum black hole evaporating into nothing. The internal
energy or related temperature of the spacetime frame fields is an
additional source of gravity, although macroscopically it is very
small, at the quantum level its contribution is necessary for a thermal
equilibrium 3-space just right being a GSRS, otherwise, the Equivalence
Principle would breakdown. In this paper, the extended quantum Equivalence
Principle plays a fundamental role as a bridge from the quantum reference
frame theory (as a statistical fields or quantum fields theory on
the base/lab spacetime) to the quantum gravity.

To sum up, the paper can be seen as an attempt to discuss the deep
relations between these three fundamental themes: the diffeomorphism
anomaly, gravity and the spacetime thermodynamics, based on the statistical
fields theory of quantum spacetime reference frame and the quantum
Equivalence Principle. In the spirit of classical general relativity,
if we trust the Equivalence Principle, one can not in principle figure
out whether one is in an absolute accelerating frame or in an absolute
gravitational background, which leads to a general covariance principle
or diffeomorphism invariance of the gravitational theory. However,
at the quantum level, the issue is a little subtle. If an observer
in an accelerating frame sees the Unruh effect, i.e. thermal particles
are creating in the ``vacuum'', which seems leading to the unitarily
inequivalence between the vacuums of, for instance, an inertial frame
and an accelerating frame, and hence the diffeomorphism invariance
is seen breakdown discussed as the anomaly in the paper. The treatment
of the anomaly in the paper is that, the anomaly is only canceled
in an observer's lab up to UV scale, where the frame can be considered
classical, rigid and cold, while at general scale the anomaly is not
completely canceled. Whether one can figure out that he/she is in
an absolute accelerating frame by detecting the anomaly (Shannon $\tilde{N}$
term) at general scale (e.g. by thermodynamic experiments detecting
the vacuum thermal particle creation and hence find the non-unitarity)?
We argue that if the answer is still ``NO!'' in the spirit of the
general relativity, the anomaly term coming from a quantum general
coordinates transformation must be also equivalently interpreted as
the effects of spacetime thermodynamics and gravity. Because the 2nd
order moment fluctuation of the quantum coordinates or a non-trivial
manifolds density $u$, which gives rise to the diffemorphism anomaly,
also contributes to other 2nd order quantities (series coefficients
at second spacetime derivative) such as (i) the acceleration (second
time derivative of coordinates, e.g. leading to uniform acceleration
expansion or other acceleration discrepancies in the universe \citep{Luo:2019iby}),
(ii) the gravity or curvature (second spacetime derivative of metric,
e.g. see (\ref{eq:dg(2)}) and (\ref{eq:density ric bound})) and
(iii) the thermal broadening (second spatial derivative of the manifolds
density or the ensemble density, e.g. see (\ref{eq:sigma=00003Dtau*g})
and (\ref{eq:tau-T})) at the same (2nd) order. In this sense, the
validity of the classical Equivalence Principle would be generalized
to the quantum level to incorporate the effects of the quantum fluctuation
of the spacetime coordinates or frame fields, so that, one in principle
still can not figure out and distinguish whether he/she is in an accelerating
frame, or in a gravitational field or in a thermal spacetime (as a
new dimension of the Equivalence Principle), these three things have
no absolute physical meanings and are indistinguishable any more in
the framework. The classical Equivalence Principle asserts the equivalence
of the first two things at the first order (mean level), the quantum
Equivalence Principle asserts the equivalence of the three things
even at the second order (variance level), even higher order.

\section*{Data availability statement}

All data that support the findings of this study are included within
the article.
\begin{acknowledgments}
This work was supported in part by the National Science Foundation
of China (NSFC) under Grant No.11205149, and the Scientific Research
Foundation of Jiangsu University for Young Scholars under Grant No.15JDG153.

\bibliographystyle{plain}

\end{acknowledgments}

\end{document}